\UseRawInputEncoding
\documentclass[preprint,prc,amsmath,amssymb,superscriptaddress,noffotinbib,preprintnumbers]{revtex4-1}
\usepackage{url,color,colordvi,epsfig,pstricks}
\usepackage{hyperref}
\usepackage{cleveref}
\usepackage{graphicx,bm,dcolumn}
\hypersetup{
    colorlinks=true, 
    linktoc=all, 
    linkcolor=blue, 
    citecolor=blue}
    
\begin{document}

\title{Recent Progress in Low Energy Neutrino Scattering Physics and Its Implications for the Standard and Beyond the Standard Model Physics}

\author{V.~Pandey}\email{vpandey@fnal.gov}
\affiliation{Fermi National Accelerator Laboratory, Batavia, Illinois 60510, USA}

\begin{abstract}
Neutrinos continue to provide a testing ground for the structure of the standard model of particle physics as well as hints towards the physics beyond the standard model. Neutrinos of energies spanning over several orders of magnitude, originating in many terrestrial and astrophysical processes, have been detected via various decay and interaction mechanisms. At MeV scales, there has been one elusive process, until a few years ago, known as coherent elastic neutrino-nucleus scattering (CEvNS) that was theoretically predicted over five decades ago but was never observed experimentally. The recent experimental observation of the CEvNS process by the COHERENT collaboration at a stopped pion neutrino source has inspired physicists across many subfields. This new way of detecting neutrinos has vital implications for nuclear physics, high-energy physics, astrophysics, and beyond. CEvNS, being a low-energy process, provides a natural window to study light, weakly-coupled, new physics in the neutrino sector. Leveraging orders of magnitude higher CEvNS cross section, new physics can be searched with relatively small detectors. \\ \\
In this review, we intend to provide the current status of low energy neutrino scattering physics and its implications for the standard and beyond the standard model physics. We discuss low energy sources of neutrinos with a focus on neutrinos from the stopped pions. Stopped pion sources cover energies in the tens of MeVs and are almost optimal for studying CEvNS. Several worldwide experimental programs have been or are being set up to detect CEvNS and new physics signals in the near future with complementary detection technologies and physics goals. We discuss the general formalism of calculating the tree-level CEvNS cross section and the estimated theoretical uncertainties on the CEvNS cross section stemming from different sources. We also discuss the inelastic scattering of tens of MeV neutrinos that have implications for supernova detection in future neutrino experiments. The stopped-pion facilities are also a near-ideal tens of MeV neutrino source to study inelastic neutrino-nucleus cross sections. We discuss how the CEvNS experiments can be used as a testing ground for the Standard Model (SM) weak physics as well as in searching for the Beyond the Standard Model (BSM) physics signals. Any deviation from the SM predicted event rate either with a change in the total event rate or with a change in the shape of the recoil spectrum, could indicate new contributions to the interaction cross-section. The SM implications include the study of weak nuclear form factor and weak mixing angle. The BSM studies include non-standard interactions, neutrino electromagnetic properties, and sterile neutrino searches. Stopped pion facilities are also a copious source of neutral and changed mesons that allow study of several dark sector physics scenarios such as vector portal models, leptophobic dark matter as well as axion-like particle searches.
\end{abstract}

\hfill\preprint{FERMILAB-PUB-23-245-ND}

\maketitle

\tableofcontents

\clearpage

\section{Introduction}\label{intro}

Neutrinos, often referred to as elusive or ghostly elementary particles, are fascinating. Starting from their postulation as a mere theoretical idea of an undetectable particle by Wolfgang Pauli in 1930, to now known as the most abundant matter particle in the Universe, neutrinos have played a prominent role in our understanding of the nature of the Universe. In recent years, neutrinos have not only provided a testing ground for the structure of the Standard Model (SM) of particle physics but also continue to provide us hints towards the physics beyond the SM. One of the most prominent one of those is the discovery of neutrino mixing and oscillation that imply that neutrinos can no longer be considered as massless particles as described in the Standard Model. SM provides the framework describing how neutrinos interact with leptons and quarks through weak interactions but it does not answer fundamental questions about neutrinos. What is the origin of neutrino mass and why are they orders of magnitude smaller compared to other SM particles? We don't know if neutrinos are Dirac particles or Majorana particles. Are there more neutrinos than three flavors, consistent with leptons, or are more of them as some experimental anomalies suggest? Neutrinos continue to provide both a testing ground for the SM and direct evidence for physics beyond the SM~\cite{Huber:2022lpm, Balantekin:2022jrq, deGouvea:2022gut, Acharya:2023swl}. \\ \\
Neutrinos originate via various mechanisms in many terrestrial and astrophysical processes, covering energies as low as from sub-eV scale to as high as EeV scale. We have detected neutrino from a variety of astrophysical (e.g., solar, supernova) and terrestrial (e.g., reactors and accelerator) sources~\cite{Davis:1968cp,Kamiokande-II:1987idp,Bionta:1987qt,IceCube:2018cha} using a variety of interaction processes ranging from inverse beta decay to scattering off quarks, nucleons, and nuclei. At MeV scale energies, which is the focus of this article, neutrinos have been detected via several distinct interaction channels including neutrino-electron elastic scattering, as well as neutral and charged current inelastic interactions on nucleons and nuclei~\cite{Formaggio:2012cpf} and via inverse-beta decay process. Among them there has been one elusive process, until a few years ago, known as Coherent Elastic Neutrino Nucleus Scattering (CEvNS) that was first postulated over nearly five decades ago.  \\ \\
CEvNS was suggested soon after the experimental discovery of the weak neutral current in neutrino interactions~\cite{Stodolsky:1966zz, Freedman:1973yd, Kopeliovich:1974mv}. In his 1974 article, Freedman suggested that {\it ``if there is a weak neutral current, then the elastic process $\nu+A \rightarrow \nu + A$ should have a sharp coherent forward peak just as the  $e+A \rightarrow e + A$ does''}~\cite{Freedman:1973yd}. Freedman went ahead and declared that the experimental detection of CEvNS would be an {\it ``act of hubris''} due to the associated {\it ``grave experimental difficulties''}. The experimental difficulty that Freedman referred to was despite the fact that the CEvNS cross section is larger due to the $\sim N^2$ enhancement it receives. The only experimental signature of the coherent elastic process is the kinetic energy $T$ of the recoiling nucleus. The maximum recoil energy is limited by the kinematics of the elastic scattering

\begin{equation}
T_{\rm max} = \frac{E_\nu}{1 + M_A/(2E_\nu)}
\end{equation}
\\
where $E_\nu$ is the incoming neutrino energy and $M_A$ is the mass of the target nuclei. For tens of MeV incident neutrino energies, where CEvNS cross section is supposed to dominate, and for medium-sized nuclei, the recoil energy amounts to several tens of keV, making it experimentally challenging to detect. \\ \\
Over after nearly four decades of its predictions by Freedman, the CEvNS signal was finally detected by the COHERENT collaboration in 2017~\cite{COHERENT:2017ipa}. The necessary keV scale low-threshold needed to detect CEvNS signal benefited from the recent developments in the detector technologies that are primarily driven by dark sector searches that also rely on tiny nuclear recoils. Typically, the recoil energy is collected in the form of scintillation photons or ionized charge, depending on the detector technology. The COHERENT collaboration announced the detection of the first CEvNS signal using a stopped--pion neutrino source at the Spallation Neutron Source (SNS) at Oak Ridge National Laboratory with a CsI[Na] scintillating crystal detector, an experimental discovery of CEvNS signal at the $6.7\sigma$ confidence level~\cite{COHERENT:2017ipa, COHERENT:2018imc}. In the following years, COHERENT collaboration presented another CEvNS measurement with a single-phase liquid argon detector~\cite{COHERENT:2019iyj,COHERENT:2020iec}, and a follow-up CsI[Na]~\cite{COHERENT:2021xmm} measurement with a larger exposure. \\ \\ 
This new way of detecting neutrinos has wider implications for border communities that span nuclear physics, particle physics, astrophysics, and beyond. Leveraging orders of magnitude higher CEvNS cross section, one could do groundbreaking searches with relatively small detectors as opposed to the typically large detector size needed for most neutrino experiments. CEvNS, being a low-energy process, provides a natural window to study light, weakly-coupled, new physics in the neutrino sector~\cite{Barranco:2005yy,Scholberg:2005qs,Barranco:2007tz,Dutta:2015nlo,Lindner:2016wff,Coloma:2017ncl, Farzan:2017xzy, Billard:2018jnl, AristizabalSierra:2018eqm, Brdar:2018qqj, Abdullah:2018ykz, AristizabalSierra:2019zmy, Miranda:2019skf,Bell:2019egg, AristizabalSierra:2019ufd, Cadeddu:2019eta, Coloma:2019mbs, Canas:2019fjw, Dutta:2019eml, Denton:2020hop, Skiba:2020msb, Cadeddu:2020nbr, Canas:2018rng}. \\ \\
The remainder of this article is organized as follows. In Sec.~\ref{sec:sources}, we discuss low-energy sources of neutrinos with a focus on neutrinos from the stopped pion sources. In Sec.~\ref{sec:CEvNS}, we lay out the general formalism of calculating the tree-level CEvNS cross section and discuss the estimated theoretical uncertainties on the CEvNS cross section stemming from different sources. Tens of MeV neutrinos also scatter via inelastic neutrino-nucleus scattering, we discuss those in Sec.~\ref{sec:inelastic}. These processes have implications for supernova detection in future neutrino experiments. The observable final-state particles of these inelastic scattering have typical energies of the same order as the incident neutrino energies. CEvNS experiments are, in principle, sensitive to inelastic processes as well if they have the dynamic range. In Sec.~\ref{sec:experimental_landscape}, we briefly review current and proposed CEvNS experimental facilities. In Sec.~\ref{sec:sm_physics}, we discuss how the CEvNS experiments can be used as a testing ground for the SM weak physics. We continue to discuss the implications of CEvNS physics for the global efforts of the BSM physics searches in Sec.~\ref{sec:bsm_physics}. CEvNS provides a natural window to study light, weakly-coupled, beyond the standard model physics in the neutrino sector. Finally, we summarize in Sec.~\ref{sec:summary}.


\section{Low Energy Neutrino Sources}\label{sec:sources}

The coherent elastic process dominates only at tens of MeV neutrino energies. The tens of MeV neutrinos come from several sources including nuclear reactors, accelerator produced decay at rest sources, as well as astrophysical sources such as supernova and solar neutrinos. Neutrinos from reactors have been detected using the inverse beta decay reaction, $\bar{\nu}_e + p \rightarrow e^+ + n$, by observing both the outgoing positron and coincident neutron. Nuclear reactors are copious sources of electron anti-neutrinos therefore, reactors have long been the sources of choice for CEvNS searches. However, since the typical reactor neutrino energies are of the order of few MeV and even if the coherence condition for the recoil is largely preserved, the scattering signal is sub-kev energy scale nuclear recoil making it even harder to detect even for many sensitive detection technologies. Furthermore, the total CEvNS cross section scales as a function of incident neutrino energy, therefore, higher energies are beneficial up until the point at which CEvNS is strongly dominated relative to inelastic scattering. Neutrinos from stopped pions sources cover energies in the tens of MeV scale and are almost optimal for studying CEvNS, finding a sweet spot where the CEvNS rate is high enough and recoil energies are more easily detectable above the threshold. So far, CEvNS is observed only at the decay at rest sources. Therefore most of the discussions in this paper focus on neutrinos from a pion decay at rest source. We comment on other low-energy neutrino sources where appropriate.


\subsection{Stopped Pion Source}

\begin{figure}
\centering
\includegraphics[width=0.7\columnwidth]{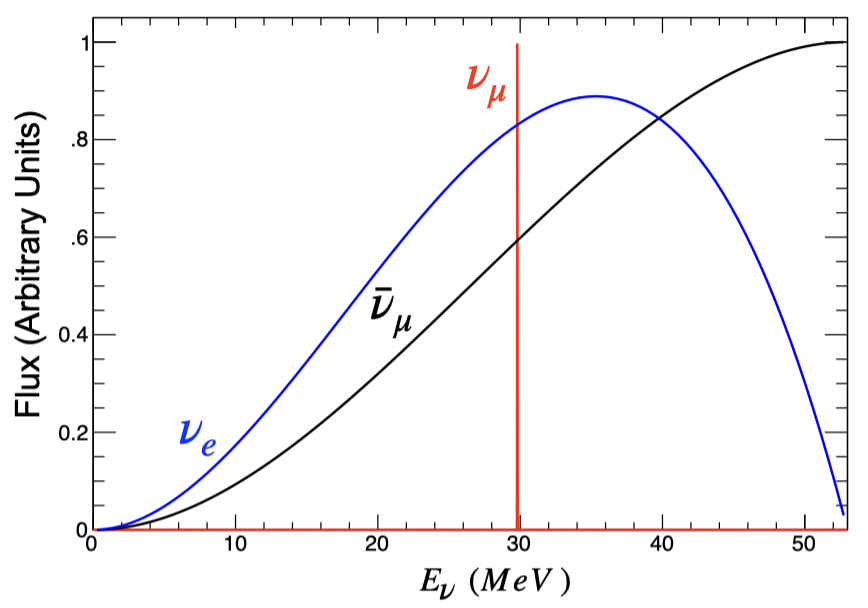}
\caption{Standard neutrino energy distribution of pion decay at rest neutrinos with a 29.8 MeV monoenergetic $\nu_\mu$, while the energies of $\nu_e$s and $\bar{\nu}_\mu$s ranges upto $m_\mu/2$.}
\label{fig:flux_1}
\end{figure}

\begin{figure}
\centering
\includegraphics[width=0.7\columnwidth]{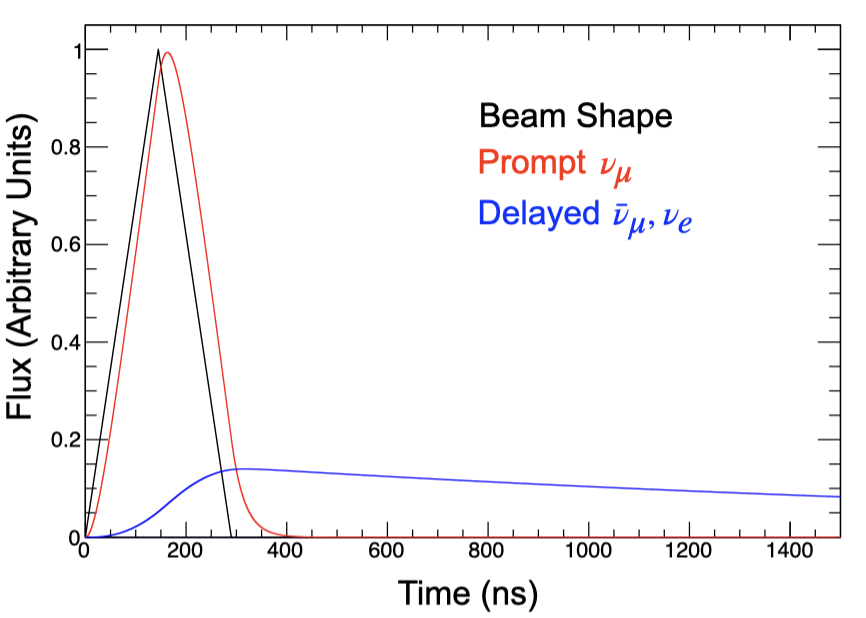}
\caption{Standard timing distribution of pion decay at rest neutrinos with a \textit{prompt} $\nu_\mu$ and \textit{delayed} $\nu_e$ and $\bar{\nu}_\mu$ signal.}
\label{fig:flux_2}
\end{figure}

An intense beam of protons accelerated to hundreds of MeV to GeV scale, directed to collide with a target, producing a copious number of secondary hadrons. Protons with energies $>$ 300 MeV will produce large numbers of pions, these pions can lose energy in dense material, stop and decay after coming to rest. Negative pions are often captured by nuclei. To produce clean stopped-pion neutrinos: (a) the optimum proton energies are of the order of 1 GeV or less~\cite{Alonso:2010fs}, this suppresses the decay-in-flight component that is heavily used in typical accelerator-based short- and long-baseline neutrino facilities, (b) the target is preferred to be dense to allow the pions to stop and decay at rest. \\ \\
The dominant neutrino production from stopped pions is from the weak-interaction two-body \textit{prompt} decay

\begin{equation}
    \pi^{+} \rightarrow \mu^+ + \nu_\mu ~~(\text{decay time:}~ \tau \sim 26 ~{\text ns})
\end{equation}
followed by a three-body \textit{delayed} decay of muons
\begin{equation}
    \mu^{+} \rightarrow e^+ + \nu_e + \bar{\nu}_\mu ~~(\text{decay time:}~ \tau \sim 2.2 ~\mu{\text s})
\end{equation}
producing a well know spectrum shape. The spectral functions are given by

\begin{equation}
     \Phi_{\nu_\mu}(E_\nu) = \frac{2m_\pi}{m_\pi^2-m_\mu^2} \delta\left(1-\frac{2E_\nu m_\pi}{m_\pi^2-m_\mu^2}\right)
\end{equation}
\begin{equation}
       \Phi_{\nu_e}(E_\nu) = \frac{192}{m_\mu}\left(\frac{E_\nu}{m_\mu}\right)^2 \left(\frac{1}{2}-\frac{E_\nu}{m_\mu}\right)
\end{equation}
\begin{equation}
  \Phi_{\bar\nu_\mu}(E_\nu) = \frac{64}{m_\mu}\left(\frac{E_\nu}{m_\mu}\right)^2 \left(\frac{3}{4}-\frac{E_\nu}{m_\mu}\right) \\ \\
\end{equation}
\\ 
For a pion decay at rest source, $E_\nu^\text{max}=m_\mu/2$ where $m_\mu = 105.65$ MeV is the muon mass. The well-known energy spectrum is shown in Fig.~\ref{fig:flux_1} with a 29.8 MeV monoenergetic $\nu_\mu$ while the energies of $\nu_e$s and $\bar{\nu}_\mu$s range upto $m_\mu/2$. Fig.~\ref{fig:flux_2} shows the standard timing distribution with a \textit{prompt} $\nu_\mu$ and \textit{delayed} $\nu_e$ and $\bar{\nu}_\mu$ signal. The pulsed time structure gives a strong handle on suppressing the background. \\ \\
There are a few percent of radiative corrections on this flux decaying from pions and muons, these are evaluated in Ref.~\cite{Tomalak:2021lif} by comparing the tree-level neutrino energy spectra with the $\mathcal{O}(\alpha)$ contributions. Radiative effects modify the expected neutrino fluxes from around the peak region by 3--4 permille.
%

\section{Coherent Elastic Neutrino Scattering off Nuclei}\label{sec:CEvNS}

The coherent elastic neutrino-nucleus scattering process occurs when a neutrino scatters off an entire nucleus, exchanging a $Z^0$ boson, transferring some of its momenta to the nucleus as a whole, but creating no internal excitations of the nucleus or ejected particles. It's \textit{elastic} in the sense that no new particles are created in the scattering and the residual nucleus stays in its ground state. For neutrinos carrying a few tens of MeV energies and scattering off medium-sized nuclei, a dominant fraction of interactions are expected to be of coherent type. 

\begin{figure}
\centering
\includegraphics[width=0.7\columnwidth]{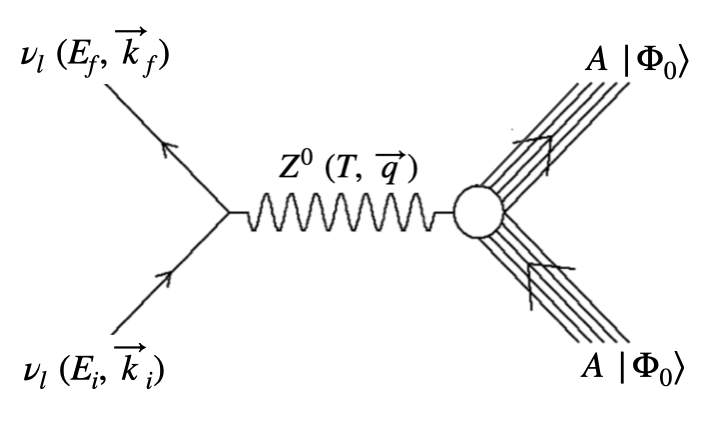}
\caption{Diagrammatic representation of the CEvNS process where a single $Z^{0}$ boson is exchanged between the neutrino and the target nucleus. The nucleus stays in its ground state and a keV scale nuclear recoil energy is deposited in the detector.}
\label{fig:cevns_diagram}
\end{figure}


\subsection{Tree-level Cross Section}

A neutrino with four momentum $k_i = (E_i,\vec{k}_i)$ scatters off the nucleus, which is initially at rest in the lab frame with $p_{A} = (M_A,\vec{0})$, exchanging a $Z^{0}$ boson. The neutrino scatters off, carrying away four momentum $k_f = (E_f,\vec{k}_f)$ while the nucleus remains in its ground state and receives a small recoil energy $T$, so that $p'_{A} = (M_A + T,\vec{p}'_{A})$ with $|\vec{p}'_{A}| = \sqrt{(M_A + T)^2 - M_A^2}$ and $T = q^2/2M_{A}$. Here, $M_A$ is the rest mass of the nucleus, $q = |\vec{q}|$ is the absolute value of the three--momentum transfer which is of the order of keV for neutrino energies of tens of MeV, $Q^2 \approx q^2 = |\vec{k}_f-\vec{k}_i|^2$, and the velocity dependent factor in the denominator refers to the relative velocity of the interacting particles. The process is schematically shown in Fig.~\ref{fig:cevns_diagram}. \\ \\
The initial elementary expression for the cross section reads

\begin{equation}
\begin{aligned}
\mathrm{d}^6\sigma &=\frac{1}{\left| \vec{v}_i - \vec{v}_A \right|}\frac{m_i}{E_i}\frac{m_f}{E_f}\frac{\mathrm{d}^3\vec{k}_f}{(2\pi)^3}\frac{M_A}{M_A + T}\frac{\mathrm{d}^3\vec{p}'_{A}}{(2\pi)^3} \\
 &\times (2\pi)^4  \overline{\sum}_{fi}\left| \mathcal{M} \right|^2 \delta^{(4)}(k_i + p_A - k_f -p'_A).
\end{aligned}
\end{equation}
\\
This expression can be integrated to yield the expression for the cross section differential in neutrino scattering angle $\theta_f$:
\begin{equation}
\begin{aligned}
\frac{\mathrm{d}\sigma}{ \mathrm{d}\cos{\theta_f}} &=\frac{m_i}{E_i}\frac{m_f}{E_f}\frac{M_A}{M_A + T} \frac{E_f^2}{2\pi}f_{rec}^{-1} \overline{\sum}_{fi}\left| \mathcal{M} \right|^2.
\end{aligned}
\end{equation}
\\
The recoil factor reads
\begin{equation}
f_{rec} = \frac{E_i}{E_f}\frac{M_A}{M_A+T}.
\end{equation}
\\
Working out the Feynman amplitude one gets

\begin{equation}
\overline{\sum}_{fi}\left| \mathcal{M} \right|^2 = \frac{G_F^2}{2}L_{\mu\nu}W^{\mu\nu},
\end{equation}
\\
with the nuclear tensor $W^{\mu\nu}$ reading

\begin{equation}
W^{\mu\nu} = \overline{\sum}_{fi} (\mathcal{J}^{\mu}_{nuc})^\dagger \mathcal{J}^{\nu}_{nuc}.
\end{equation}
\\
The summation symbols in these expressions denote summing and averaging over initial and final polarizations, respectively. The nuclear tensor depends on the nuclear current transition amplitudes:

\begin{equation}
\mathcal{J}^{\mu}_{nuc} = \langle \Phi_\textrm{0} | \widehat{J}^\mu(\vec{q}) | \Phi_\textrm{0} \rangle .
\end{equation}
\\
Under the assumption that the nuclei of interest are spherically symmetric with $J^\pi = 0^+$ and taking the z--axis to be along the direction of $\vec{q}$, one only needs to take into account the zeroth and third component of the nuclear current's vector part, which are furthermore connected through vector current conservation (CVC):
\begin{equation}
q^\mu \widehat{J}_\mu(\vec{q}) = 0. \\
\end{equation}
Through performing the necessary algebra, one arrives at the final expression

\begin{equation}
\frac{\mathrm{d}\sigma}{ \mathrm{d}\cos{\theta_f}} = \frac{G_F^2}{2\pi} \frac{E_f^3}{E_i} \left[\frac{Q^4}{q^4}(1+\cos{\theta_f}) |\mathcal{J}^V_0|^2\right]
\end{equation}
\\
where $\mathcal{J}^V_0$ is the transition amplitude induced by the nuclear current. One can then safely approximate $\frac{Q^4}{q^4} \approx 1$ and express the differential cross section as a function of the neutrino scattering angle $\theta_f$ as:

\begin{equation}\label{Eq:xs_angular}
\frac{\mathrm{d}\sigma}{ \mathrm{d}\cos{\theta_f}} = \frac{G_F^2}{2\pi}  \frac{E_f^3}{E_i} (1+\cos{\theta_f}) \frac{Q_W^2}{4}F_{W}^2(Q^2)
\end{equation}
\\
where $G_F$ is the Fermi coupling constant, and $Q_W$ the tree-level weak nuclear charge:

\begin{equation}\label{eq:weakcharge}
Q^{2}_{W} = [g_p^V Z+g_n^V N]^2 = [(1-4\sin^2\theta_\text{W}) Z-N]^2
\end{equation}
\\
with coupling constants $g_n^V = -1$ and $g_p^V = (1-4\sin^2\theta_\text{W})$. $N$ and $Z$ are the nucleus' neutron and proton number, and $\theta_W$ is the weak mixing angle. The value is such that $\sin^2{\theta_W} = 0.23857$, which is valid at low momentum transfers~\cite{Ishikawa:2018rlv}. \\ \\
Here we have introduced the elastic form factor, $F_{W}^2(Q^2)$, which we will discuss later in this subsection. In elastic scattering, the entire nuclear dynamics is encoded in this form factor. Equivalently one can express the differential cross section as a function of the nuclear recoil $T$, which reads:

\begin{equation}\label{Eq:cevns_xs}
\frac{\mathrm{d}\sigma}{ \mathrm{d}T} = \frac{G^{2}_{F}}{\pi} M_{A} \left(1-\frac{T}{E_{i}}-\frac{M_A T}{2 E^2_i}\right)~\frac{Q^2_{W}}{4}~F_{W}^2(Q^2).
\end{equation}
\\
In Eq.~(\ref{Eq:xs_angular}) and~(\ref{Eq:cevns_xs}), we have expressed the CEvNS kinematic distribution both in neutrino scattering angle, $\theta_f$, and in nuclear recoil energy $T$. $T=q^2/(2M)=E_\nu-E_\nu'$ is the nuclear recoil energy (taking values in $[0,2E_\nu^2/(M+2E_\nu)]$). Terms of order $T/E_\nu \lesssim 2E_\nu/M_A$ are usually neglected since they will be negligible for neutrino energies $E_\nu \lesssim 50$ MeV accessible at the stopped pion sources. The cross section represents the truly ``coherent'' contribution, in the sense that the nuclear structure physics that enter the definition of weak form factor $F_\text{W}$, indeed scale with $Z$ and $N$. \\ \\
In most experiments, the only signal of a CEvNS event is a nuclear recoil energy deposition. In principle, future experiments with more advanced detector technologies may be able to detect both nuclear recoil and angular distribution simultaneously. Such capabilities are already being explored in some dark-matter experiments and will significantly enhance the physics capabilities of future CEvNS experiments~\cite{Abdullah:2020iiv}. The cross section can also be expressed in terms of the direction of the recoil, converting the recoil to an angular spectrum. This is referred to in the literature as the Directional Recoil Spectrum (DRS)~\cite{Abdullah:2020iiv} where the angles are those of the scattered nucleus measured with respect to the incident neutrino direction and can be written as

\begin{equation}
\frac{\text{d}^2R}{\text{d}\Omega\text{d}T}=\frac{1}{2\pi}
\left .\frac{\text{d}\sigma}{\text{d}T}\right|_{E_\nu=\varepsilon}\,
\frac{\varepsilon^2}{E_\nu^\text{min}}
\left .\frac{\text{d}\Phi}{\text{d}E_\nu}\right|_{E_\nu=\varepsilon}\,,
\end{equation}
\\
where $\text{d}\Phi/\text{d}E_\nu$ is the differential neutrino flux, $E_\nu^\text{min} = \sqrt{M T/2}$, and 

\begin{equation}
\frac{1}{\varepsilon}=\frac{\cos \theta}
{E_\nu^\text{min}}-\frac{1}{M}\,.
\end{equation}
\\
To switch variables directly between $T$ and $\Omega$ one can use the following relation and the associated Jacobian:

\begin{equation}
T=\frac{2ME_\nu^2\cos^2\theta}{(E_\nu+M)^2-E_\nu^2\cos^2\theta}\,.
\end{equation}
\\
The directional and energy double differential cross section can be written by noting that the scattering has azimuthal symmetry about the incoming neutrino direction. Integrating over outgoing nuclear recoil energy gives

\begin{equation}
\label{dsigmadOmega}
\frac{\text{d}\sigma}{\text{d}\Omega} = \frac{G_F^2}{16\pi^2}
Q_\text{W}^2
E_\nu (1+\cos \theta)
\big[F_\text{W}(q^2)\big]^2 \,, 
\end{equation}
\\
where the angle is defined as
\begin{equation}
\text{d}\Omega = 2\pi \cos \theta \text{d}\theta
\end{equation}
and $\theta$ is the scattering angle between the direction of the incoming and outgoing neutrino.  \\ \\
The scattering process' cross section is proportional to the squared magnitude of the transition amplitude induced by the nuclear current. Since the relevant ground state to ground state transition for spherically symmetrical nuclei is $0^+ \rightarrow 0^+$, only the vector part of the current will contribute. The amplitude can be expressed as

\begin{equation}\label{Eq:cevnsformfactor1}
\begin{aligned}
 \mathcal{J}^V_0 &= \langle \Phi_0 | \widehat{J}_0^V(\vec{q}) | \Phi_0 \rangle \\
 &= \int e^{i\vec{q}\cdot \vec{r}} \langle \Phi_0 | \widehat{J}_0^V(\vec{r}) | \Phi_0 \rangle \\
 &= \frac{1}{2}\left[\left(1 - 4 \sin^2{\theta_W} \right) f_p(\vec{q})F_{p}(Q^2) \right. \\
 &- \left. f_n(\vec{q})F_{n}(Q^2) \right],
\end{aligned}
\end{equation}
\\
where we have inserted the impulse approximation (IA) expression for the nuclear current, as a sum of single--body operators:

\begin{equation}\label{Eq:cevnscurrent}
\widehat{J}_0^V(\vec{r}) = \sum_i F^Z(Q^2,i)\delta^{(3)}(\vec{r}-\vec{r}_i), 
\end{equation}
with
\begin{equation}\label{Eq:cevnscurrent2}
\begin{aligned}
F^Z(Q^2,i) &= \left( \frac{1}{2}-\sin^2{\theta_W} \right)(F_{p} - F_{n})\tau_3(i) \\
&- \sin^2{\theta_W}(F_{p} + F_{n}),
\end{aligned}
\end{equation}
\\
where we used the convention $\tau_3(i)=+1$ for proton, -1 for neutrons.
Furthermore, $f_p(\vec{q})$ and $f_n(\vec{q})$ are the Fourier transforms of the proton and neutron densities, respectively. $F_{p}$ and $F_{n}$ are proton and neutron form factors, for which we adopt the standard Galster parametrization. Note that using a more sophisticated parametrization of the form factor, other than Galster, will not affect the results at the energies relevant to this work.
 The overall structure of the transition amplitude consists of products of the weak charge with two factors: the nuclear form factor, determined by the spatial distribution of the nucleons in the nucleus, as well as the nucleon form factor. We arrive at the expression:
 
\begin{equation}\label{Eq:cevnsformfactor2}
\begin{aligned}
 F_{W}(Q^2) &= \frac{1}{Q_W}\left[\left(1 - 4 \sin^2{\theta_W} \right) f_p(\vec{q})F_{p}(Q^2) \right. \\
  & \left. -  f_n(\vec{q})F_{n}(Q^2) \right] = \frac{2}{Q_W}\mathcal{J}^V_0,
 \end{aligned}
\end{equation}
\\
such that the form factor becomes 1 in the static limit. Note that in writing down the functional dependence, we can make use of the non--relativistic approximation $Q \approx |\vec{q}|$, valid in the energy regime considered.  
 

\subsection{Uncertainty on the Cross Section}

At tree-level, the theoretical uncertainty on the CEvNS cross section is driven by the uncertainty on the weak form factor of the nucleus. Although, in deriving CEvNS cross section in the previous section, a number of subtleties have been ignored that including subleading kinematic effects, axial-vector contributions and radiative corrections. In this subsection,  we will first discuss the uncertainty on the tree-level cross section driven by weak form-factor and then briefly discuss other subleading uncertainties. \\ \\
The CEvNS cross section is proportional to the weak form factor of Eq.~(\ref{Eq:cevnsformfactor2}). In general, the form factor can be reasonably approximated by several different functional forms. The simplest way is to denote neutron and proton form factors in the Eq.~(\ref{Eq:cevnsformfactor2})
%
%
as Fourier transforms of neutron and proton densities considering the nucleus to be spherically symmetric. 

\begin{equation}\label{Eq:Fn}
F_n(Q^2) = \frac{4\pi}{N} \int dr~r^2~\frac{\sin(Qr)}{Qr}~\rho_n(r)
\end{equation}

\begin{equation}\label{Eq:Fp}
F_p(Q^2) = \frac{4\pi}{Z} \int dr~r^2~\frac{\sin(Qr)}{Qr}~\rho_p(r)
\end{equation}
\\
where $\rho_n(r)$ and $\rho_p(r)$ are neutron and proton density distributions normalized to the neutron and proton numbers. The value of the nuclear form factors in the limit for q $\rightarrow$ 0 is 1. A small value of the coefficient of the proton form factor in Eq.~(\ref{Eq:cevnsformfactor2}) makes the weak form factor and hence CEvNS is mainly sensitive to neutron density distribution.  The charge density of a nucleus is strongly dominated by the protons and has been extensively studied with impressive precision in elastic electron scattering experiments started in the late 1950’s~\cite{Hofstadter:1956qs} followed by subsequent refinements over the decades~\cite{DeVries:1987atn, Fricke:1995zz, Angeli:2013epw}. On the other hand, the neutron density distributions are hard to determine, and various efforts using hadronic probes were plagued by uncontrolled model--dependent uncertainties associated with the strong interaction~\cite{Thiel:2019tkm}. Electroweak processes such as parity-violating electron scattering (PVES)~\cite{Donnelly:1989qs} and CEvNS have long been considered clean and model-independent probes for extracting ground-state neutron densities. Both of these, though long considered experimentally challenging, are becoming a reality in recent years. \\ \\
Phenomenological form factors, such as Helm~\cite{Helm:1956} and Klein-Nystrand~\cite{KN:1999}, are widely used in the CEvNS community where density distributions are represented by analytical expressions. The empirical value of proton rms radius, extracted from elastic electron scattering data, is often used to evaluate the proton form factor, and the same parameterization (or a variation of that) is assumed for the neutron form factor.\\ \\
In the Helm approach~\cite{Helm:1956}, the nucleonic density distribution is described as a  convolution of a uniform density with radius $R_0$ and a Gaussian profile characterized by the folding width $s$, accounting for the surface thickness and the form factor is expressed as:

\begin{equation}
F_{\text{Helm}}(q^2) = \frac{3 j_1(qR_0)}{qR_0} e^{-q^2s^2/2}
\end{equation}
\\
where $j_1(x) = \sin(x)/x^2 - \cos(x)/x$ is a spherical Bessel function of the first kind. $R_0$ is an effective nuclear radius given as: $R_0^2 = (1.23 A^{1/3} - 0.6)^2 + \frac{7}{3} \pi^2 r_0^2 - 5 s^2$ with $r_0$ = 0.52 fm and $s$ = 0.9 fm, fitted~\cite{Duda:2006uk, Lewin:1995rx} to muon spectroscopy and electron scattering data compiled in~\cite{Fricke:1995zz}. \\ \\
The Klein--Nystrand (KN) form factor, adapted by the COHERENT Collaboration, is obtained from the convolution of a short-range Yukawa potential with range $a_k$ = 0.7 fm over a Woods--Saxon distribution approximated as a hard sphere with radius $R_A = 1.23 A^{1/3}$ fm~\cite{KN:1999}. The resulting form factor is expressed as:
\begin{figure*}
\centering
\includegraphics[width=0.7\textwidth]{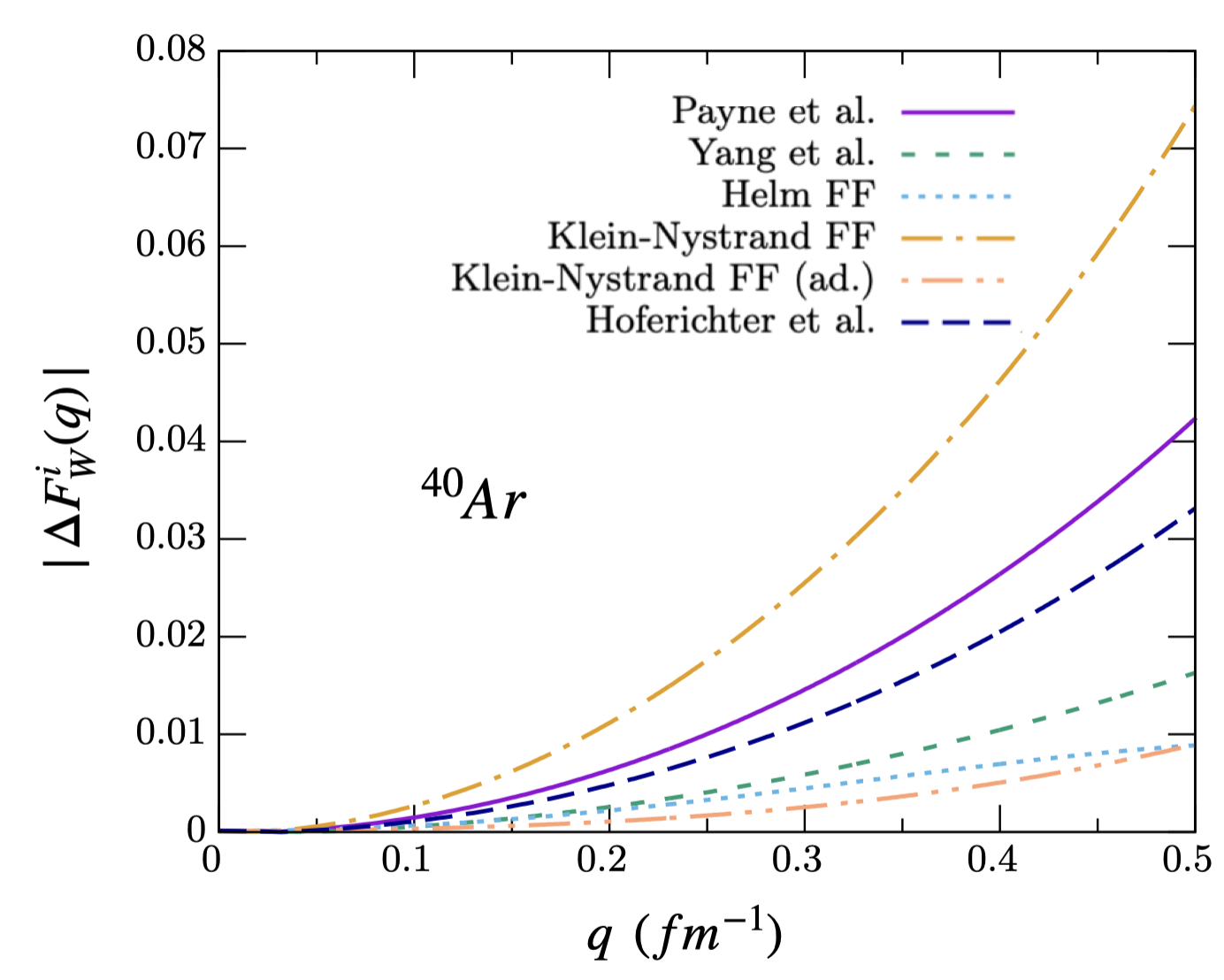}
\caption{Relative differences in the $^{40}$Ar weak form factor predictions of Payne {\it et al.}~\cite{Payne:2019wvy}, Yang {\it et al.}~\cite{Yang:2019pbx}, Hoferichter {\it et al.}~\cite{Hoferichter:2020osn}, Helm~\cite{Helm:1956}, Klein--Nystrand~\cite{KN:1999} and the adapted Klein--Nystrand~\cite{AristizabalSierra:2019zmy, Papoulias:2019xaw}, all with respect to HF calculations of Van Dessel {\it et al.}~\cite{VanDessel:2020epd}. Figure adapted from Ref.~\cite{VanDessel:2020epd}.}
\label{Fig:arcomps}
\end{figure*}

\begin{equation}
F_{\text{KN}}(q^2) = \frac{3 j_1(qR_A)}{qR_A} \left[\frac{1}{1+q^2a_k^2} \right].
\end{equation}
\\
An adapted version of the KN form factor, (ad.) KN form factor, is often used where $R_A$ is defined as $R_A = \sqrt{\frac{5}{3}r_0^2 - 10 a_k^2}$ utilizing measured proton rms radii $r_0$ of the nucleus~\cite{AristizabalSierra:2019zmy, Papoulias:2019xaw}. The measured proton rms radii of $^{40}$Ar is,  $r_0 = 3.427$ fm~\cite{Angeli:2013epw}. \\ \\
More involved nuclear structure calculations which describe a more accurate picture of the nuclear ground state such as first-principles calculation using coupled--cluster theory from first principles of Payne {\it et al.}~\cite{Payne:2019wvy}, shell-model calculations of Hoferichter {\it et al.}~\cite{Hoferichter:2020osn} where form factors are calculated using a large--scale nuclear shell model, relativistic mean--field method of Yang {\it et al.}~\cite{Yang:2019pbx} where form factors predictions are informed by  properties of finite nuclei and neutron star matter, and Hartree--Fock approach of Van Dessel {\it et al.}~\cite{VanDessel:2020epd} where form factors are computed in a mean-field using Skyrme potential, has been reported in recent years. In order to quantify differences between different form factors and the CEvNS cross section due to different underlying nuclear structure details, we can consider quantities that emphasize the relative differences between the results of different calculations, arbitrarily using Hartree--Fock (HF) as a reference calculation, as follows:

\begin{figure*}
\centering
\includegraphics[width=0.7\textwidth]{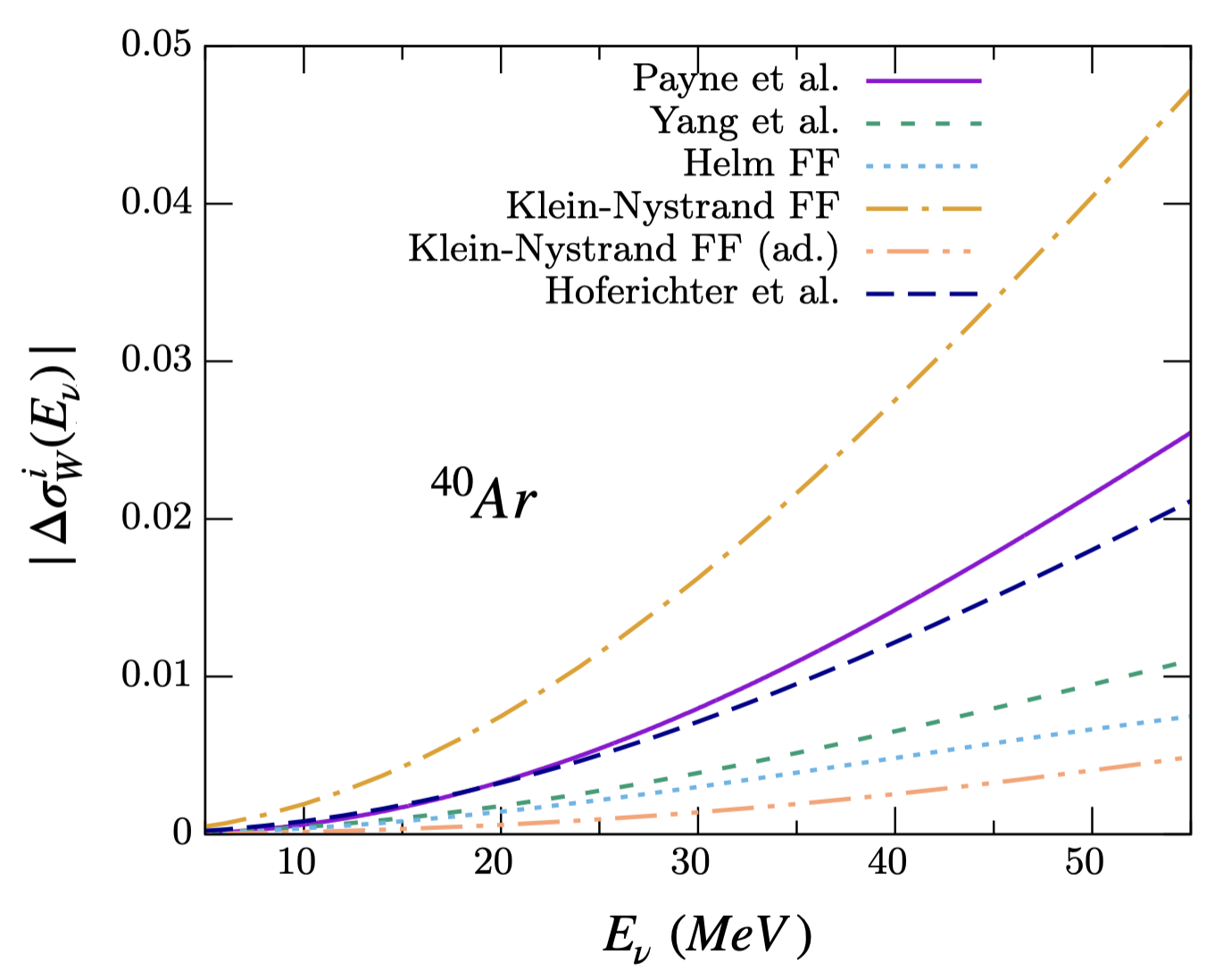}
\caption{Relative differences in the $^{40}$Ar CEvNS cross section predictions of Payne {\it et al.}~\cite{Payne:2019wvy}, Yang {\it et al.}~\cite{Yang:2019pbx}, Hoferichter {\it et al.}~\cite{Hoferichter:2020osn}, Helm~\cite{Helm:1956}, Klein--Nystrand~\cite{KN:1999} and the adapted Klein--Nystrand~\cite{AristizabalSierra:2019zmy, Papoulias:2019xaw}, all with respect to HF calculations of Van Dessel {\it et al.}~\cite{VanDessel:2020epd}. Figure adapted from Ref.~\cite{VanDessel:2020epd}.}
\label{Fig:arcomps_2}
\end{figure*}
\begin{equation}
|\Delta F_{\text W}^{i}(q)|~=~ \frac{|F_{\text W}^{i}(q) - F_{\text W}^{\text {HF}}(q)|}{|F_{\text W}^{\text {HF}}(q)|},
\end{equation}

\begin{equation}
|\Delta \sigma_{\text W}^i(E_\nu)|~=~ \frac{|\sigma_{\text W}^{i}(E_\nu) - \sigma_{\text W}^{\text {HF}}(E_\nu)|}{|\sigma_{\text W}^{\text {HF}}(E_\nu)|},
\end{equation}
\\
where $i$ refers to calculations from different approaches as discussed above.\\ \\
The relative differences are shown in Fig.~\ref{Fig:arcomps} and Fig.~\ref{Fig:arcomps_2}. We show only the low--momentum part of the weak form factor to a maximum value of $q$ = 0.5 fm$^{-1}$ ($\sim$ 100 MeV) that corresponds to a maximum incoming neutrino energy of E $\sim$ 50 MeV. The relative differences are shown on a linear scale. At smaller energies, the momentum transfer is low and hence the differences between form factors are also small. For higher energies, the available momentum transfer increases and therefore, the differences between the form factors become more prevalent. The differences in model predictions amount to $< 7.5\%$ over the entire momentum transfer range. The differences rise rapidly at the higher end of the $q$ range. This translates into relative differences in CE$\nu$NS cross sections, $\Delta \sigma(E)$, of $< 5\%$ over the whole energy range, where $E \lesssim 55$ MeV, relevant for neutrinos from pion decay-at-rest. \\ \\
In writing down the CEvNS cross section, Eq.~(\ref{Eq:cevns_xs}), only the vector operators were considered. In principle, the axial-vector operator adds an additional contribution that is not coherently enhanced, including this, modifies the cross section to the form

\begin{equation}
\label{CEvNS_SM}
\frac{\text{d} \sigma}{\text{d} T}=\frac{G_F^2 M}{4\pi}\bigg(1-\frac{M T}{2E_\nu^2}-\frac{T}{E_\nu}\bigg)Q_\text{W}^2\big[F_\text{W}(q^2)\big]^2
+\frac{G_F^2M}{4\pi}\bigg(1+\frac{M T}{2E_\nu^2}-\frac{T}{E_\nu}\bigg)F_A(q^2)\,,
\end{equation}
\\
with an axial-vector form factor $F_A(q^2)$~\cite{Hoferichter:2020osn}. The axial-vector form factor depends on the axial charges and radii of the nucleon. This contribution vanishes for spin-zero nuclei such as $^{40}$Ar. \\ \\
The CEvNS cross section expression of Eq.~(\ref{Eq:cevns_xs}) holds true at tree-level, in which case $Q_\text{W}$ are flavor universal and apply both to neutrino and electron scattering. Once including radiative corrections, process- and flavor-dependent contributions arise, in such a way that separate weak charges need to be defined. Electrons and muons running in loops introduce a non-trivial dependence on the momentum transfer due to their relatively light masses. These break the flavor universality because of mass-dependent electromagnetic radiative corrections. For CEvNS, the corresponding radiative corrections have been studied in Ref.~\cite{Tomalak:2020zfh}. At next-to-leading order (NLO) in the electromagnetic coupling constant $\alpha$, photon-mediated scattering takes place and the cross section inherits a flavor-dependent contribution entering with a charge form factor of the nucleus. 

\begin{equation}
    \frac{\mathrm{d} \sigma_{\nu_\ell}}{\mathrm{d} T} = \frac{\mathrm{G}_\mathrm{F}^2 M_\mathrm{A}}{4\pi} \left( 1-\frac{T}{E_\nu} -\frac{M_\mathrm{A} T}{2E_\nu^2} \right) \left( \mathrm{F}_\mathrm{W} \left(Q^2\right)  + \frac{\alpha}{\pi} [\delta^{\nu_\ell} + \delta^\text{QCD}]   \mathrm{F}_\mathrm{ch}(Q^2)\right)^2,
\end{equation}
\\ 
The expression depends on the weak, $\mathrm{F}_\mathrm{W}$, and charge, $\mathrm{F}_\mathrm{ch}$, nuclear form factors. The charge form factor enters multiplied by $\delta^{\nu_\ell}$ and $\delta^\text{QCD}$ which are radiative corrections. The corrections induced by hadronic and/or quark loops, proportional to $\delta^\text{QCD}$, are flavor independent, whereas the corrections from charged leptons, proportional to $\delta^{\nu_\ell}$, depend on the neutrino flavor $\ell$.\\ \\
A detailed total theoretical uncertainty on the CEvNS cross sections $^{40}$Ar nucleus was estimated Ref.~\cite{Tomalak:2020zfh}, and is shown in Tab.~\ref{tab:errors_Argon}. The estimated error budget accounts for uncertainties stemming from a variety of sources including nuclear, nucleon, and quark levels. At higher energies, the main source of uncertainty for the CEvNS cross section comes from nuclear physics. In fact, this can be traced down to the error of the neutron distribution inside the nucleus. 
\begin{table}[t]
\centering
\footnotesize
\centering 
\begin{tabular}{|c|c|c|}   
\hline          
Incident Neutrino Energy  &  Nuclear Level Uncertainty on $^{40} \mathrm{Ar}$ & Total Estimated Theoretical Uncertainty on $^{40} \mathrm{Ar}$ \\
\hline
10 (\text{MeV}) & 0.04\% & 0.58\% \\
30 (\text{MeV}) &1.5\%  & 1.65\% \\
50 (\text{MeV}) & 4.0\%  & 4.05\%  \\
\hline
\end{tabular}
\caption{Estimated theoretical errors budget on the CEvNS cross section on on$~^{40} \mathrm{Ar}$ target. Table adapted from Ref.~\cite{Tomalak:2020zfh}. 
\label{tab:errors_Argon} }
\end{table}
%


\subsection{Input from Parity Violating Electron Scattering}\label{CEvNS_PVES}

CEvNS and Parity Violating Electron Scattering (PVES) are intimately connected to each other. From the formal point of view, both processes are described in first order perturbation theory via the exchange of an electroweak gauge boson between a lepton and a nucleus. While in CEvNS the lepton is a neutrino and a $Z^0$ boson is exchanged, in PVES the lepton is an electron, but measuring the asymmetry allows one to select the interference between the $\gamma$ and $Z^0$ exchange. As a result, both the CEvNS cross section  and the PVES asymmetry depend on the weak form factor $F_W(Q^2)$, which  is mostly determined by the neutron distribution within the nucleus.  The latter builds an even stronger anchor between CEvNS and PVES. \\ \\
The key experimental observable in the elastic scattering of longitudinally polarized electrons from the unpolarized spin-0 nucleus is the parity-violating asymmetry ${A}_\mathrm{PV}$. The parity-violating asymmetry arises from the interference of $\gamma$-mediated and $Z$-mediated scattering diagrams. 
The asymmetry ${A}_\mathrm{PV}$ is determined from the fractional difference in cross sections between the scattering of positive and negative helicity electrons

\begin{equation}
    A_{pv}=\frac{d\sigma/d\Omega_+-d\sigma/d\Omega_-}{d\sigma/d\Omega_++d\sigma/d\Omega_-}
    \label{eq.Apv}
\end{equation}
\\
where $\pm$ refers to the polarization of the electron. In the Born approximation at low momentum transfer, $\mathrm{A}_\mathrm{PV}$ is proportional to the ratio of the weak to the charge form factors of the nucleus

\begin{equation}
    A_{pv}=\frac{G_Fq^2|Q_W|}{4\pi\alpha\sqrt{2}Z}\frac{F_W(q^2)}{F_{ch}(q^2)}.
    \label{eq.Apv_2}
\end{equation}
\\
For a given nucleus, if $\mathrm{F}_\text{ch}(Q^2)$ is already known from the elastic electron scattering experiment, one can extract $\mathrm{F}_\text{W}(Q^2)$ from measured $\mathrm{A}_\mathrm{PV}$ in at the momentum transfer of the experiment after accounting for radiative corrections and Coulomb distortion effects not considered in the Born approximation~\cite{Horowitz:1999fk}. Coulomb distortions can be theoretically calculated by solving the Dirac equation for an electron moving in a nuclear potential~\cite{Yennie:1954zz,Yennie:1965zz,Kim:1996ua,Kim:2001sq} and are relatively well understood~\cite{Horowitz:1998vv}. \\ \\
The PREX experiment at the Jefferson Lab (JLab) has recently provided the first model-independent determination of the weak-charge form factor of $^{208}$Pb at $\mathrm{F}_\text{W}(\langle Q^2\rangle ) = 0.204\pm0.028$ at the average momentum transfer of the experiment $\langle Q^2\rangle \approx 8800~\mathrm{MeV}^2$~\cite{Abrahamyan:2012gp, Horowitz:2012tj}. The follow-up PREX-II experiment is underway to improve the precision of that measurement. Another PVES experiment at JLab, CREX, is planned to measure the weak-charge form factor of $^{48}$Ca~\cite{Kumar:2020ejz}. In practice, however, both PREX-II and CREX measurements will make weak form factor measurements on a single value of the momentum transfer and is not expected to perform measurements at several values of the momentum transfer. Future facilities such as the MESA facility in Mainz, envisioned to start operations in a few years, will also be suited for high-precision parity-violating experiments~\cite{Becker:2018ggl}. Tab.~\ref{tab:parityviolation} summarizes current and near-future PVES experiments. It is worth noting that CEvNS can be used to probe the weak form factor only at low momentum transfers where the process remains coherent, but accesses a continuum of four-momentum transfers. In contrast, PVES experiments are usually carried out at a single value of the momentum transfer at a time. A combination of measurements from these two independent and complementary scattering techniques is ideal since systematic uncertainties are largely uncorrelated. This will then provide an empirical extraction of a nucleus' weak form factor in a clean and model-independent fashion. \\ \\
\begin{table}[tb]
\centering
\begin{tabular}{ | c | c c c c |}
\hline
Experiment    & Target& $q^2$ (GeV$^2$)  & $A_{pv}$ (ppm)     & $\pm\delta R_n$ (\%) \\
\hline
PREX at JLab & $^{208}$Pb & 0.00616 & $0.550\pm0.018$ & 1.3 \\
CREX at JLab & $^{48}$Ca & 0.0297 & & 0.7\\
Qweak at JLab & $^{27}$Al & 0.0236 & $2.16\pm 0.19$ & 4\\
MREX at MESA & $^{208}$Pb & 0.0073 & & 0.52\\
\hline
\end{tabular}
\caption{\label{tab:parityviolation}Parity violating elastic electron scattering experiments.}
\end{table} 
In principle, parity-violating electron scattering experiments offer the least model-dependent and most precise approach to experimentally probing the neutron distribution. Any result that will come from the PVES program with the goal of pinning down the neutron-skin thickness will help improve our understanding of the weak form factor and hence influence  CEvNS. However, CEvNS  has also been proposed as an alternative and attractive opportunity in the future to constrain the neutron distribution and the neutron radius in nuclei~\cite{Amanik:2009zz,Patton:2012jr,Cadeddu:2017etk}, provided that enough statistics can be reached. \\ \\ 
The main difference lies in the choice of the nuclear target, which is determined by practical considerations. In the case of PVES, the targets need to be stable (or almost stable) neutron-rich nuclei, such as $^{208}$Pb and $^{48}$Ca, that do not present low-lying excited states that would contribute to the background noise.  In the case of CEvNS, isotopes of sodium, argon, germanium,  cesium and iodine will be used, as the low cost allows to build large detectors with these materials. Because various electroweak observable correlate~\cite{Yang:2019pbx} with each other, theoretical calculations will help to further connect the various nuclear targets and the two endeavors of CEvNS and PVES.  For example, we can expect that constraints  experimentally determined on the neutron-skin thickness of one nuclear target will affect the prediction of the weak form factor of another target. CEvNS experiments also prefer detector materials with low scintillation or ionization thresholds in order to efficiently measure low-energy nuclear recoils. Quite the contrary is needed as target material in parity violation electron scattering experiments: in this case, the highest the excited state of the nucleus, the lower the contamination of the elastic asymmetries by inelastic contributions from the excited state. In addition, due to the high intensity of the electron beam, a high melting temperature of the target material is also desirable.


\section{Inelastic Neutrino Scattering off Nuclei}\label{sec:inelastic}

\begin{figure}
\centering
\includegraphics[width=0.7\columnwidth]{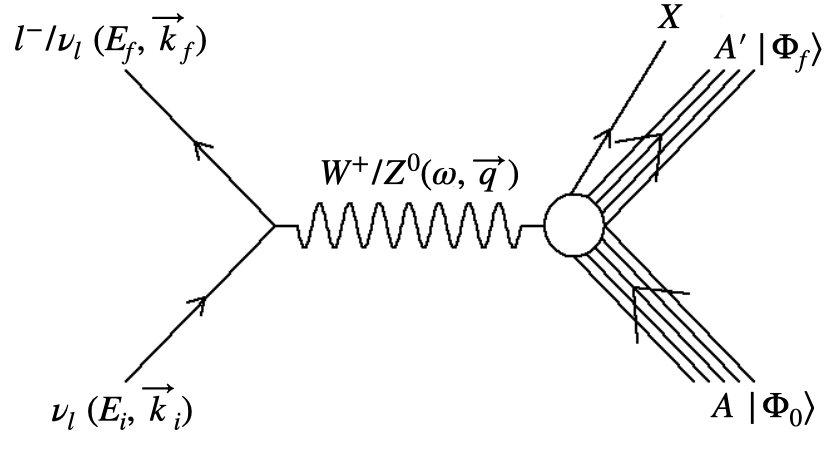}
\caption{Diagrammatic representation of the inelastic neutrino-nucleus scattering where a single $W^{+}$ (CC) or $Z^{0}$ (NC) boson is exchanged between neutrino and target nucleus, exciting the nucleus into low-lying nuclear states, followed by nuclear de-excitation products, $X$ (gamma or a nucleon), with energies of the same order as the incident neutrino energies.}
\label{fig:inelastic}
\end{figure}
\begin{figure}
\centering
\includegraphics[width=0.7\columnwidth]{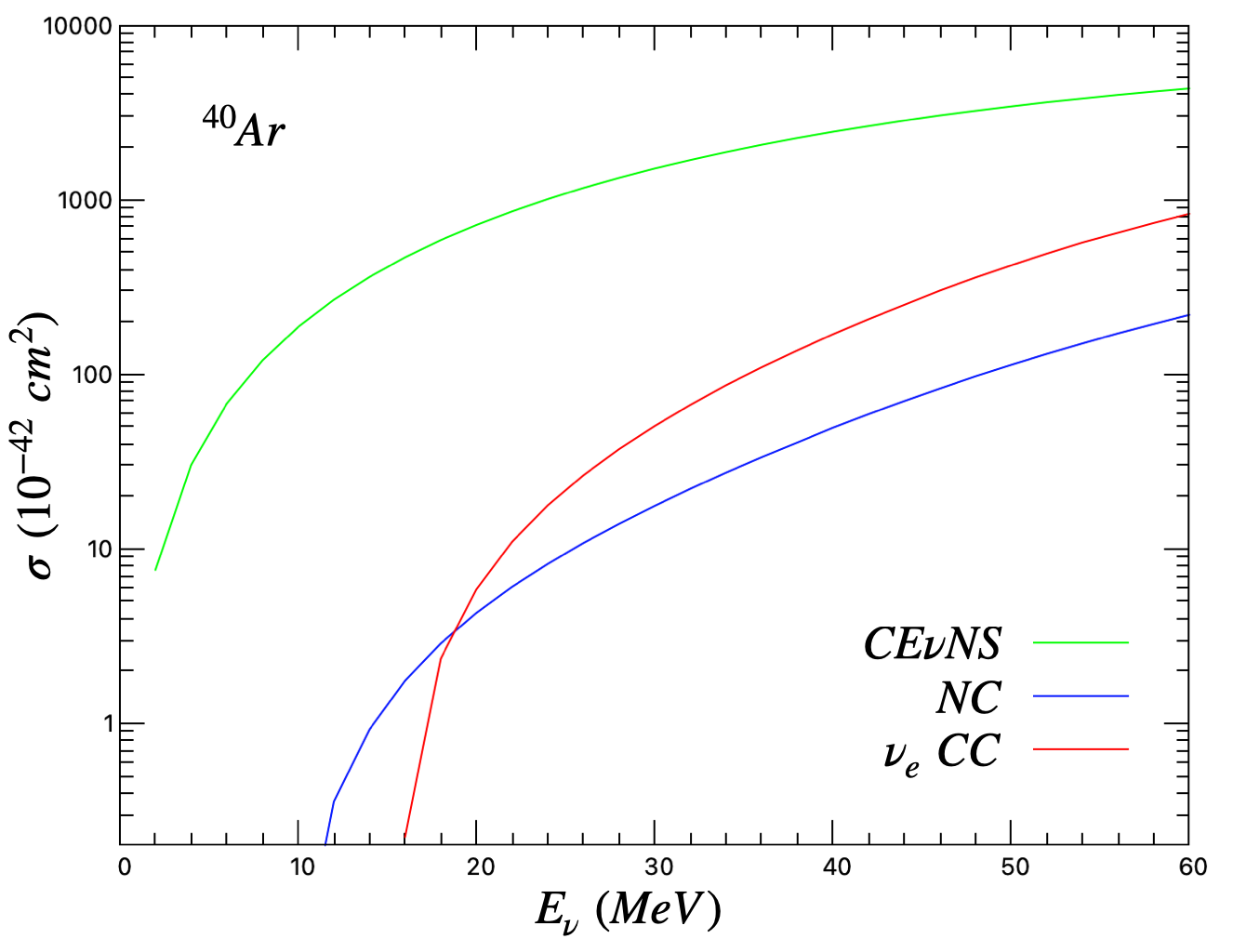}
\caption{CEvNS cross section compared with CC and NC inelastic scattering cross section on argon.}
\label{fig:xsec}
\end{figure}

CEvNS experiments at stopped--pion sources are also sensitive to inelastic neutrino-nucleus interactions. For neutrino energies less than about $\sim$100 MeV, the CEvNS interaction channel dominates the neutrino-nucleus cross section over inelastic charged-current (CC) and neutral-current (NC) neutrino-nucleus interactions. In the inelastic NC or CC scattering, shown in Fig.~\ref{fig:inelastic}, the neutrino excites the target nucleus to a low-lying nuclear state, followed by nuclear de-excitation products such as gamma rays or ejected nucleon. The interaction cross sections for these processes lack the $N^2$ enhancement associated with CEvNS and, therefore, tend to be at least one order of magnitude smaller than that of CEvNS process, as shown in Fig.~\ref{fig:xsec}. The observable final-state particles of these inelastic scattering have typical energies of the same order as the incident neutrino energies. \\ \\
The inelastic neutrino-nucleus scattering process is schematically shown in Fig.~\ref{fig:inelastic}. A neutrino with four-momentum $k_i = (E_i,\vec{k}_i)$ scatters off the nucleus, which is initially at rest in the lab frame, exchanging a $W^+$ (CC) or a $Z^{0}$ (NC) boson. The nucleus receives four momentum $Q = (\omega, \vec{q})$, where $\omega = E_i - E_f$ and $\vec{q} = \vec{k}_i - \vec{k}_f$, while the scattered lepton carries away four momentum $k_f = (E_f,\vec{k}_f)$. For an inclusive process, the hadronic part of the final states is integrated out. The inelastic neutrino-nucleus differential cross section of this process can be written as 

\begin{equation}\label{eq:xsec}
\begin{aligned}
\frac{\mathrm{d}^3\sigma}{\mathrm{d}\omega\mathrm{d}\Omega} =& \sigma_W E_f k_f \zeta^2(Z',E_f) \\
&\times \left( v^{\mathcal{M}} R^{\mathcal{M}} + v^{\mathcal{L}} R^{\mathcal{L}} + v^{\mathcal{ML}} R^{\mathcal{ML}} \right.  \\
& + \left. v^{T} R^{T} + h v^{TT} R^{TT} \right),
\end{aligned}
\end{equation}
\\
with the Mott-like cross section prefactor $\sigma_W$ defined as

\begin{equation*}
\sigma_W^{CC} = \left(\frac{G_F \cos{\theta_c}}{2\pi} \right)^2, ~\sigma_W^{NC} = \left(\frac{G_F}{2\pi} \right)^2,
\end{equation*}
\\
where $G_F$ is the Fermi constant and $\cos{\theta_c}$ the Cabibbo angle. The factor $\zeta^2(Z',E_f)$ is introduced in order to take into account the distortion of the scattered lepton wave function in the Coulomb field of the final nucleus with $Z'$ protons, in the case of CC interaction~\cite{VanDessel:2019obk, Pandey:2014tza}. In the NC case $\zeta^2(Z,E_f)$ equals $1$. The influence of the lepton helicity on the cross section is encoded in $h$ which is $+$ for neutrinos and $-$ for antineutrinos. \\ \\
The $v$--factors are leptonic functions that are entirely determined by lepton kinematics. The $R$--factors are the nuclear response functions that depend on the energy and momentum transfer ($\omega$, $q$) and contain all the nuclear information involved in this process. The indices $L$ and $T$ correspond to longitudinal and transverse contributions, relative to the direction of the momentum transfer. The leptonic coefficients $v^{\mathcal{M}}$, $v^{\mathcal{L}}$, $v^{\mathcal{M\mathcal{L}}}$, $v^{T}$, and $v^{TT}$ are given as

\begin{figure}
\centering
\includegraphics[width=0.7\columnwidth]{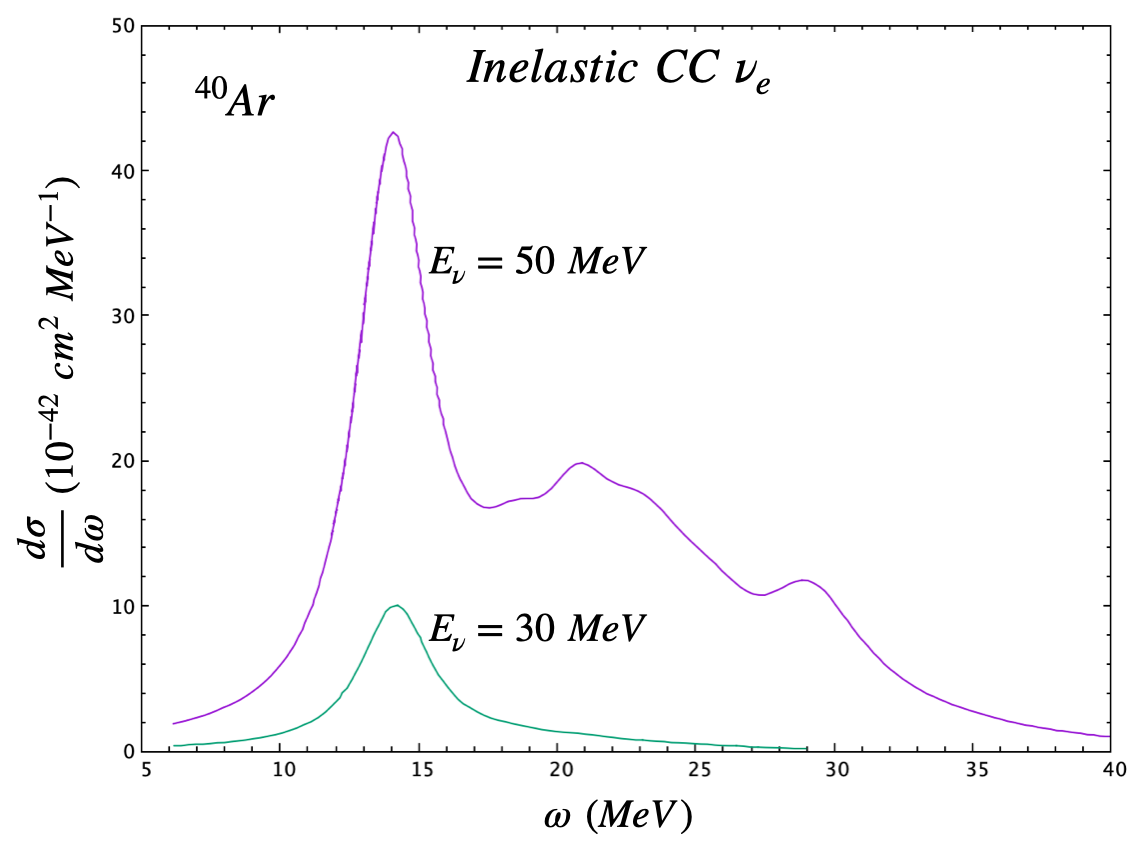}
\caption{Charged current neutrino-argon differential cross section for neutrino energy 30 and 50 MeV shown as a function of the energy transferred to the nucleus ($\omega$).}
\label{fig:xse_2}
\end{figure}
\begin{eqnarray}
 v^{\mathcal{M}} = \left[1+\frac{\kappa_f}{\varepsilon_f}\cos\theta\right], 
\end{eqnarray}
\begin{eqnarray}
 v^{\mathcal{L}} = \left[1+\frac{\kappa_f}{\varepsilon_f}\cos\theta- \frac{2 \varepsilon_i \varepsilon_{f}}
 {|\vec{q}|^2} {\left(\frac{\kappa_f}{\varepsilon_f}\right)}^2 \sin^{2}\theta\right], 
\end{eqnarray}
\begin{eqnarray}
 v^{\mathcal{M}\mathcal{L}} = \left[\frac{\omega}{|\vec{q}|}\left(1+\frac{\kappa_f}{\varepsilon_f}\cos\theta\right)+ 
 \frac{m_{l}^{2}}{\varepsilon_{f}|\vec{q}|}\right], 
\end{eqnarray}
\begin{eqnarray}
 v^{T} = \left[1-\frac{\kappa_f}{\varepsilon_f}\cos\theta+\frac{\varepsilon_i \varepsilon_{f}}{|\vec{q}|^2} 
 {\left(\frac{\kappa_f}{\varepsilon_f}\right)}^2 \sin^{2}\theta\right],  
\end{eqnarray}
 \begin{eqnarray}
 v^{TT} = \left[\frac{\varepsilon_i+\varepsilon_{f}}{|\vec{q}|}\left(1-\frac{\kappa_f}{\varepsilon_f}\cos\theta\right)-
 \frac{m_{l}^{2}}{\varepsilon_{f}|\vec{q}|}\right], 
\end{eqnarray}
\\
and response functions $R^{\mathcal{M}}$,  $R^{\mathcal{L}}$,  $R^{\mathcal{ML}}$,  $R^{T}$, 
and $R^{TT}$ are defined as

\begin{eqnarray}
 R^{\mathcal{M}} = |\langle J_f|| \widehat{\mathcal{M}}_J(|\vec{q}|)|| J_i \rangle|^2, 
\end{eqnarray}
\begin{eqnarray}
 R^{\mathcal{L}} = |\langle J_f|| \widehat{\mathcal{L}}_J(|\vec{q}|)|| J_i \rangle|^2, 
\end{eqnarray}
\begin{eqnarray}
 R^{\mathcal{ML}} = ~\mathcal{R}\left[\langle J_f|| \widehat{\mathcal{L}}_J(|\vec{q}|)|| J_i \rangle 
 \langle J_f|| \widehat{\mathcal{M}}_J(|\vec{q}|)|| J_i \rangle^{\ast} \right],
\end{eqnarray}
\begin{eqnarray}
 R^{T} = \left[ |\langle J_f|| \widehat{\mathcal{J}}_J^{mag}(|\vec{q}|)|| J_i \rangle|^2 + 
 |\langle J_f|| \widehat{\mathcal{J}}_J^{el}(|\vec{q}|)|| J_i \rangle|^2 \right], \nonumber \\ 
\end{eqnarray}
\begin{eqnarray}
 R^{TT} = ~\mathcal{R}\left[\langle J_f|| \widehat{\mathcal{J}}_J^{mag}(|\vec{q}|)|| J_i \rangle 
 \langle J_f|| \widehat{\mathcal{J}}_J^{el}(|\vec{q}|)|| J_i \rangle^{\ast} \right]. \nonumber \\ 
\end{eqnarray}
Here $\widehat{\mathcal{M}}_J$, $\widehat{\mathcal{L}}_J$, $\widehat{\mathcal{J}}_J^{mag}$ and $\widehat{\mathcal{J}}_J^{el}$
are the Coulomb, longitudinal, transverse  magnetic, and transverse electric operators, respectively~\cite{OConnell:1972edu, Walecka:1995}.\\ \\
\begin{figure}
\centering
\includegraphics[width=0.7\columnwidth]{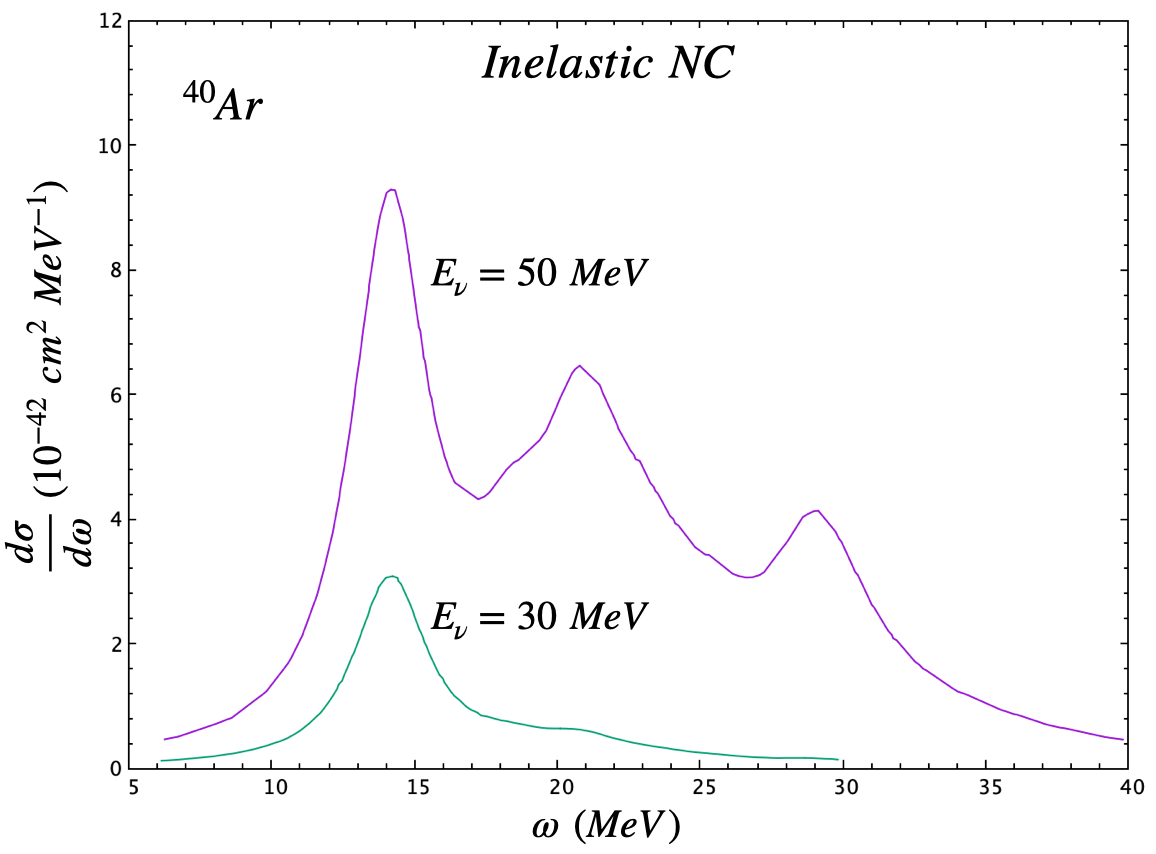}
\caption{Neutral current neutrino-argon differential cross section for neutrino energy 30 and 50 MeV shown as a function of the energy transferred to the nucleus ($\omega$).}
\label{fig:xse_3}
\end{figure}
The nuclear responses are function of the transition amplitude, ${J}_\mu^{nucl}(\omega,q)$, between the initial $| \Phi_\textrm{0} \rangle$ and final $| \Phi_\textrm{f} \rangle$ state:

\begin{equation}\label{eq:current}
{J}_\mu^{nucl}(\omega,q) = \langle \Phi_\textrm{f} | \hat{J}_\mu(q) | \Phi_\textrm{0} \rangle,
\end{equation}
\\
where the nuclear current, $\hat{J}_\mu({q})$, is the Fourier transform of the nuclear current operator in coordinate space:

\begin{equation}
\hat{J}_\mu(q) = \int \mathrm{d} {x} e^{i{x}\cdot{q}} \hat{J}_\mu({x}).
\end{equation}
\\
Nuclear responses are computed within a nuclear model. Fig.~\ref{fig:xse_2} and~\ref{fig:xse_3} show inelastic CC and NC cross-section on $^{40}$Ar as a function of $\omega$ for incoming neutrino energy of 30 and 50 MeV, calculated within a microscopic many-body nuclear theory approach of Refs.~\cite{Jachowicz:2002rr, Pandey:2014tza, Pandey:2016jju, VanDessel:2019atx}. With the stopped-pion flux, only $\nu_e$ CC interactions are accessible, given that $\nu_\mu$ and $\bar{\nu}_\mu$ are below the CC threshold of $\sim$ 110 MeV, needed to create a muon. While NC interactions are available for all neutrino types, $\nu_e$, $\nu_\mu$ and $\bar{\nu}_\mu$ at the stopped-pion facility. The experimental requirements for CEvNS and inelastic signals are quite different. Larger masses are needed for inelastics, as well as the dynamic range to record MeV-scale energy depositions, while very low thresholds are not required. \\ \\
\begin{table*}
\begin{center}
\begin{tabular}{|l|l|l|l|}
\hline
  Reaction Channel & Experiment & Measurement ($10^{-42}$ cm$^2$) \\
 \hline
 $^{12}$C($\nu_e,e^-$)$^{12}$N$_{\rm g.s.}$ & KARMEN & $9.1 \pm 0.5 {\rm(stat)} \pm 0.8{\rm(sys)}$ \\
 &  E225 & $10.5 \pm 1.0 {\rm(stat)} \pm 1.0{\rm(sys)}$\\
 & LSND & $8.9 \pm 0.3 {\rm(stat)} \pm 0.9{\rm(sys)}$ \\
 & & \\
 $^{12}$C($\nu_e,e^-$)$^{12}$N$^*$ &  KARMEN & $5.1 \pm 0.6 {\rm(stat)} \pm 0.5{\rm(sys)}$ \\
 &  E225  & $3.6 \pm 2.0 {\rm(tot)}$\\
 &  LSND & $4.3 \pm 0.4 {\rm(stat)} \pm 0.6{\rm(sys)}$ \\
 & & \\
 $^{12}$C($\nu_\mu,\nu_\mu$)$^{12}$C$^*$ & KARMEN& $3.2 \pm 0.5 {\rm(stat)} \pm 0.4{\rm(sys)}$\\
 $^{12}$C($\nu,\nu$)$^{12}$C$^*$ & KARMEN & $10.5 \pm 1.0 {\rm(stat)} \pm 0.9{\rm(sys)}$ \\
\hline
 $^{56}$Fe($\nu_e,e^-$) $^{56}$Co & KARMEN & $256 \pm 108 {\rm(stat)} \pm 43{\rm(sys)}$  \\
\hline
$^{127}$I($\nu_e,e^-$)$^{127}$Xe & LSND & $284 \pm 91 {\rm(stat)} \pm 25{\rm(sys)}$ \\
$^{127}$I($\nu_e,e^-$)X & COHERENT & $920^{+2.1}_{-1.8} $ \\
\hline
$^{nat}$Pb($\nu_e,Xn$) & COHERENT & -- -- \\
\hline
\end{tabular}
\caption{Flux-averaged cross-sections measured at stopped pion facilties  on various nuclei. Experimental data gathered from the LAMPF~\cite{Willis:1980pj}, KARMEN~\cite{KARMEN:1998xmo, KARMEN:1991vkr, Maschuw:1998qh, Zeitnitz:1994kz}, E225~\cite{Krakauer:1991rf}, LSND~\cite{LSND:2001fbw, LSND:2002oco, Distel:2002ch}, and COHERENT~\cite{COHERENT:2023ffx, COHERENT:2022eoh} experiments. Table adapted from the Ref.~\cite{Formaggio:2012cpf}.} 
\label{tab:12C}
\end{center}
\end{table*}
The detection of the burst of 10s of MeV neutrinos from the galactic core-collapse supernova is one of the primary physics goals of the future DUNE experiment~\cite{DUNE:2020zfm, DUNE:2023rtr}, as stated in the DUNE’s TDR~\cite{DUNE:2020lwj, DUNE:2020ypp, DUNE:2020mra, DUNE:2020txw}
, “Detect and measure the nue flux from a core-collapse supernova within our galaxy, should one occur during the lifetime of the DUNE experiment. Such a measurement would provide a wealth of unique information about the early stages of core collapse, and could even signal the birth of a black hole". Detecting supernova will provide unique insight into the properties of neutrinos, as well as into the astrophysics of core-collapse supernova. DUNE’s capabilities of supernova neutrino detection in the relevant tens-of-MeV neutrino energy range as well as the physics to be learned from a DUNE supernova burst detection will be limited by the lack of knowledge of the inelastic neutrino-argon cross section. The inelastic neutrino-argon cross sections in this energy range have never been measured. In the absence of experimental data, the uncertainties in the theoretical calculations are not quantified at all. The theory predictions, in fact, differ by orders of magnitude, see e.g. Fig.~6 in Ref.~\cite{DUNE:2023rtr}.  In order to reconstruct the energy of the incoming neutrinos from a supernova, the energy of all final state particles needs to be known. These will include nuclear de-excitation products such as $\gamma$-rays and potential nuclear fragments (neutrons, protons, deuterons, etc). In a recent analysis performed by the DUNE collaboration, Ref. ~\cite{DUNE:2023rtr},  reports that the total inelastic neutrino-argon cross section needs to be known at about 5\% (in the absence of any external constraints) for a measurement of the integrated neutrino luminosity with less than 10\% bias with DUNE. \\ \\
The well-understood stopped-pion neutrino spectrum is a near-ideal 10s of MeV neutrino source~\cite{Scholberg:2012id} which can provide a unique opportunity to measure neutrino-argon cross sections in this energy regime.  Inelastic interactions of neutrinos with nuclei are still poorly understood:  theory is sparse and experiments have large error bars.
There are very few existing measurements,  none at better than the 10\% uncertainty level, they are summarized in Table~\ref{tab:12C}~\cite{Formaggio:2012cpf}. So far, there are no measurements on the argon nucleus performed to date. Because inelastic neutrino interactions have big uncertainties, in the future it will be crucial to measure inelastic electron scattering cross sections at energies below the 50 MeV mark and use those data to calibrate theoretical models for the neutrino scattering process. Theoretical understanding of these processes is also relatively poor, due to the strong dependence of the interaction rates on the specific initial- and final-state nuclear wavefunctions. Inelastic neutrino-argon cross sections shown in Figs.~\ref{fig:xsec}, Figs.~\ref{fig:xse_2} and Figs.~\ref{fig:xse_3} have never been measured before. CEvNS experiments at decay at rest sources, such as COHERENT~\cite{COHERENT:2020iec} and Coherent CAPTAIN-Mills~\cite{CCM} experiments are well suited to make those measurements. The technical challenge is that the experiment has to have a dynamic range of detecting keV energy recoil (signal for CEvNS), and MeV energy nuclear deexcitation and nuclear fragment products (signal for inelastic scattering) in the same detector. COHERENT experiment has recently demonstrated their capabilities of measuring inelastic cross section by performing two measurements, $^{127}$I($\nu_e,e^-$)X  and $^{nat}$Pb($\nu_e, Xn$) ~\cite{COHERENT:2023ffx, COHERENT:2022eoh}.


\section{Experimental Landscape}\label{sec:experimental_landscape}
\begin{table*}[htbp]
\centering
\begin{tabular}{|c|c|c|c|c|c|}
\hline
Experiment & Nuclear Target  & Detector Technology  & Mass (kg) & Distance from source (m) & Dates \\
\hline 
COHERENT & CsI[Na] & Scintillating crystal & 14  & 19.6 & 2015-2019 \\ 
(ORNL) & Pb, Fe & Liquid scintillator & 1,000 & 19.0  &  2015- \\
	   & NaI[Tl] & Scintillating crystal & 185 & 21.0  & 2016- \\
	   & LAr   & Noble scintillator    & 24  & 27.5  & 2017-  \\
	   & D$_2$O & Cherenkov            & 600 &  22.0   & 2022- \\	
	   & Ge     & HPGe PPC             & 18  & 21.0    & 2022- \\
	   & NaI[Tl] & Scintillating crystal & 3,388 & 24.0  & 2022- \\
\hline
CCM & LAr & Noble scintillator & 10,000  & 23.0   & 2019 -   \\ 
(LANL) &  &  &   &   &   \\ 
\hline
\end{tabular}
\caption{Current CEvNS experiments at the stopped pion sources.}
\label{tab:CEvNS_1}
\end{table*}
Several experimental programs have been or are being set up to detect CEvNS and BSM signals in the near future using stopped--pion neutrino sources as well as with reactor sources. It all started with the COHERENT collaboration reporting on the first detection of the CEvNS process in 2017. The measurement was performed with an exposure of 14.6-308 kg-days, the COHERENT collaboration identified nuclear recoil events from CEvNS viewed by a single photomultiplier tube (PMT). The measurement was well in excess of the expected background events for this exposure. A likelihood analysis considering the signal and background shapes in time and PE yielded a result of 134 $\pm$ 22 CEvNS events, with the uncertainty being primarily statistical, the SM prediction for this analysis is 178 $\pm$ 43 CEvNS events. The observed event rate was consistent with the SM prediction within uncertainties~\cite{COHERENT:2017ipa}. This led to a flurry of proposals and experiments worldwide with complementary detection technologies and physics goals. In Table~\ref{tab:CEvNS_1}, we list currently running CEvNS experiments at stopped pion sources. There are several proposed experiments at existing and planned facilities that are not included in the table but we discuss them below. For the sake of completeness, we also list CEvNS experiments at reactors in Table~\ref{tab:CEvNS_2}. These include CONNIE~\cite{CONNIE}, MINER~\cite{MINER}, $\nu$GEN~\cite{vGEN}, NUCLEUS~\cite{NUCLEUS}, RICOCHET~\cite{RICOCHET}, TEXONO~\cite{TEXONO}, NEON~\cite{NEON} and vIOLETA~\cite{vIOLETA} experiments. The current theme of reactor experiments is the observation of neutrino-nucleus elastic scattering at the kinematic regime where complete quantum-mechanical coherency is expected.  \\ \\
SNS at ORNL: The Spallation Neutron Source at the Oak Ridge National Laboratory has the most ambitious CEvNS-based experimental program~\cite{Barbeau:2021exu,Asaadi:2022ojm}. SNS is consistently running at 1 GeV proton energy and 1.4 MW beam power. By 2024 after the next round of upgrades, it will be running with 1.3 GeV proton energy and 2 MW beam power. The SNS First Target Station (FTS) proton beam consists of a linear $H^-$ ion accelerator, an accumulator ring, and a proton target. The proton target employs liquid mercury contained inside a double-walled stainless steel vessel~\cite{Henderson:2014paa, Haines:2014kna}. The SNS generates 400-nanosecond bursts of protons on target at 60 Hz frequency allowing for a highly effective suppression of backgrounds and simultaneous measurement of neutrino signal and backgrounds. A second target station (STS) with a solid tungsten target is planned for the SNS. For this stage, the total beam power will be increased to 2.8 MW and the proton beam will be split between two targets with 45 Hz to the first target and 15 Hz to the second, creating even more favorable conditions to suppress steady-state backgrounds. The COHERENT collaboration continued pursuing several additional detector technologies for CEvNS, to span a range of N values, as well as detectors to address additional physics goals.\\ \\ 
Lujan at LANL: The Lujan Center at the Los Alamos National Laboratory is a prolific source of neutrinos from decays of stopped pions and muons created by an 800 MeV proton beam.  An 800MeV protons are delivered at a rate of 20Hz in a 280 ns triangular pulse from the LANSCE beamline and interact in a thick tungsten target, copiously producing charged and neutral mesons. A 10 ton liquid argon scintillation detector, Coherent CAPTAIN-Mills (CCM), is currently operating. The CCM upright cylindrical cryostat 2.58 m in diameter and 2.25 m high. A ton-scale mass and a keV-range energy threshold allow the CCM detector to possess leading sensitivity to potential dark-sector physics signals~\cite{CCM,CCM:2021yzc,CCM:2021lhc}. \\ \\
\begin{table}[tb]
\centering
\begin{tabular}{ | c | c | c | c | }
 \hline
  Experiment & Detector Technology & Location & Source \\
  \hline
  CONNIE &Si CCDs&Brazil & Reactor \\
  CONUS& HPGe&Germany& Reactor\\
  MINER&Ge/Si cryogenic& USA& Reactor\\
  NuCleus& Cryogenic CaWO$_4$, Al$_2$O$_3$ calorimeter array& Europe& Reactor\\
  $\nu$GEN &Ge PPC&Russia & Reactor\\
  RED-100 & LXe dual phase& Russia& Reactor \\
  Riochet & Ge,Zn & France & Reactor\\
  TEXONO &p-PCGe& Taiwan& Reactor \\
  NCC-1701 &p-PCGe& Germany& Reactor \\
\hline    
\end{tabular}
\caption{A list of reactor based CEvNS experiments.}
\label{tab:CEvNS_2}
\end{table} 
JSNS2 at JPARC: The Japan Spallation Neutron Source of J-PARC is featured by a 1 MW beam of 3 GeV protons incident on a mercury target, creating an intense neutrino flux from the stopped-pion and stopped-muon decays. The JSNS2 (J-PARC Sterile Neutrino Search at J-PARC Spallation Neutron Source) experiment aims to search for the existence of neutrino oscillations and to offer the ultimate test of the LSND anomaly at a 17-ton fiducial volume Gd-dopped liquid scintillation detector, new detector is being planned to study not only CEvNS but potential low-mass dark-matter signals~\cite{Ajimura:2017fld, Ajimura:2020qni}. \\ \\
ESS: The European Spallation Source (ESS), sited in Sweden, will combine the world’s most powerful superconducting proton linac with an advanced hydrogen moderator, generating the most intense neutron beams for multi-disciplinary science. It will also generate the largest pulsed neutrino flux suitable for the detection of CEvNS. The ESS aims to achieve the power of 5 MW and proton energy of 2 GeV. Several detector technologies sensitive to keV energy nuclear recoils are being considered, these include a cryogenic undoped CsI scintillator array, silicon Charge Coupled Devices (CCDs), and high-pressure gaseous xenon detectors~\cite{Baxter:2019mcx}. \\ \\
PIP2-BD at FNAL: The Proton Improvement Project II (PIP-II) is the first phase of a major transformation of the accelerator complex underway at Fermilab to prepare the lab to host the Deep Underground Neutrino Experiment (DUNE). The completion of the PIP-II superconducting LINAC at Fermilab as a proton driver for DUNE/LBNF in the late 2020s creates an attractive opportunity to build such a dedicated beam dump facility at Fermilab, this will require the addition of an accumulator ring to bunch the PIP-II beam current into short proton pulses. A unique feature of this Fermilab beam dump facility is that it can be optimized from the ground up for HEP. Thus, relative to spallation neutron facilities dedicated to neutron physics and optimized for neutron production operating at a similar proton beam power, a HEP-dedicated beam dump facility would allow for better sensitivity to various physics goals. The facility could also accommodate multiple, 100-ton-scale HEP experiments located at different distances from the beam dump and at different angles with respect to the beam~\cite{Toups:2022yxs}.


\section{Implications for the Standard Model Physics}\label{sec:sm_physics}

Since the uncertainty on the SM predicted CEvNS cross sections is relatively small, CEvNS cross section measurement allows testing of SM weak physics. The experiments measure the number of events $N$ generated by neutrinos of a given flavor $\alpha$ and collected by a detector with a finite threshold

\begin{equation}
 \frac{dN}{dT} = N_t \sum_{\alpha=\nu_e, \nu_{\mu}, \bar{\nu}_{\mu}} \int^{E_\nu^\text{max}}_{E_\nu^\text{min}} \Phi_{\alpha}(E_\nu)~\frac{d\sigma}{dT} ~dE_\nu \label{rate}
\end{equation}
\\
where $N_t$ is a normalization constant that depends on the number of protons on target, the neutrino yield per proton, the mass of the detector, detection efficiency and the distance of the detector from the source. Any deviation from the SM predicted event rate of Eq.~({\ref{rate}), either with a change in the total event rate or with a change in the shape of the recoil spectrum, could indicate new contributions to the interaction cross-section. More generally, what can be probed is the weak nuclear form factor of a nucleus and weak mixing angle (see, Eq.~(\ref{Eq:cevns_xs})). There are many important results that can be extracted from the CEvNS measurements, recent work has considered CEvNS as a percent-level probe of SM physics~\cite{Scholberg:2005qs, Miranda:2019skf, Cadeddu:2019eta, Canas:2018rng, Papoulias:2019xaw, Baxter:2019mcx, Huang:2019ene, Bernabeu:2002nw, Bernabeu:2002pd, Papavassiliou:2005cs, Cadeddu:2018dux}. Note that for a given stopped pion production facility, experiments have control over choosing the baseline (the distance from the source to the detector), the angular placement of the detector with respect to the beam axis, and on the nuclear target employed in the detector. These can be exploited to increase the sensitivity of the primary physics goal of the experiment. An additional advantage of the stopped pion source is that one could exploit both timing and energy data. The timing profile, See Fig.~\ref{fig:flux_2}, allows the separation of the prompt neutrino flavor from the delayed neutrino flavor. 


\subsection{Weak Nuclear Form Factor}

The modest loss of coherence at stopped-pion energies can be valuable for the understanding of the nuclear structure, given that one can probe the form factor for a given nucleus as a function of $Q$. A precise measurement of the CEvNS cross section can be used to extract the weak form factor, using Eq.~(\ref{Eq:cevns_xs}), given one measures recoil spectrum shape. Observed recoil energy T can be used to determine Q; therefore, the observed CEvNS recoil energy spectrum allows one to map the effect of the weak form factor of the nucleus at low momentum transfer. \\ \\
As discussed in Sec.~\ref{sec:CEvNS}, the weak nuclear charge is strongly dominated by its neutron content. The observation of CEvNS can, therefore, further provide important nuclear structure information through the determination of the weak form factor, which constrains the neutron density distribution, at least at low momentum transfers where the process remains coherent~\cite{AristizabalSierra:2019zmy, Payne:2019wvy, Hoferichter:2020osn, Yang:2019pbx, VanDessel:2020epd, Patton:2012jr, Cadeddu:2017etk, Co:2020gwl, Ciuffoli:2018qem, Papoulias:2019lfi}. Furthermore, since proton density distributions are generally well understood, a measure of the mean radius of the neutron distribution (the “neutron radius”) enables the determination of the “neutron skin” of a nucleus — the difference between the larger neutron radius and the proton radius. These measurements complement PVES experiments not only due to 
additional data but also due to different energy ranges and nuclear targets, which could be used to calibrate nuclear-structure calculations. Furthermore, improved measurements of the neutron skin would have important consequences for the equation of the state of neutron-rich matter, which plays an essential role in understanding the structure and evolution of neutron stars~\cite{Fattoyev:2017jql, Reed:2021nqk, Lattimer:2012xj,Hebeler:2013nza,Hagen:2015yea}. With more ambitious precision measurements, axial-vector contributions to the weak nuclear response can also be determined, in principle, for nuclei with non-zero spin. \\ \\
However, arguably one of the most intricate aspects of nuclear-structure input concerns searches for physics beyond the SM. In principle, CEvNS cross sections provide constraints on the combination of nuclear responses and the BSM effects. Therefore external independent experimental information for the neutron responses, such as from the  PVES experiment would be vital. In fact, in order to derive BSM constraints beyond the level at which current nuclear-structure calculations constrain the neutron distribution, a combined analysis of multiple targets and momentum transfers is required to distinguish between nuclear structure and potential BSM contributions~\cite{Abdullah:2022zue}. 


\subsection{Weak Mixing Angle}

In quantum field theory, the weak mixing angle, $\theta_W$, depends on the energy scale at which it is
measured. There exists an experimental anomaly with respect to the SM predictions for neutrino-nucleon scattering at the $Q\sim$~GeV/c scale~\cite{NuTeV:2001whx}. Since CEvNS cross section and, therefore, the event rate depends on the weak mixing angle, Eq.~(\ref{Eq:cevns_xs}) and Eq.~(\ref{eq:weakcharge}), the measured CEvNS event counts can be used to infer the weak mixing angle. A change in $\theta_W$ will result in eventual event rate scaling. \\ \\ 
Furthermore, measurements on multiple nuclear targets will further enhance the sensitivity to extracting weak mixing angles from weak nuclear charge~\cite{Scholberg:2005qs, Miranda:2019skf, Canas:2018rng, Papoulias:2019xaw, Baxter:2019mcx, Huang:2019ene}. As CEvNS measurements become more precise in the near-future, one could extract the weak mixing angle at $Q$ values of a few tens of MeV/c; that will be competitive with other methods for determining $\theta_W$ at low $Q$ from parity-violating electron-proton scattering~\cite{Qweak:2018tjf}, Moller scattering~\cite{MOLLER:2014iki} and atomic parity violation~\cite{Roberts:2014bka}. 


\section{Implications for Beyond the Standard Model Physics}\label{sec:bsm_physics}

CEvNS, being a low-energy process, provides a natural window to study light, weakly-coupled, beyond the standard model physics in the neutrino sector. Several extensions of the SM can be explored at low energy~\cite{Barranco:2005yy, Scholberg:2005qs,  Barranco:2007tz, Lindner:2016wff, Coloma:2017ncl, Farzan:2017xzy, AristizabalSierra:2018eqm, Brdar:2018qqj, Abdullah:2018ykz, AristizabalSierra:2019zmy, Miranda:2019skf, Bell:2019egg, AristizabalSierra:2019ufd, Cadeddu:2019eta, Coloma:2019mbs, Canas:2019fjw, Dutta:2019eml, Denton:2020hop, Skiba:2020msb, Cadeddu:2020nbr, Abdullah:2020iiv, Papoulias:2019xaw, Baxter:2019mcx, AristizabalSierra:2017joc, Giunti:2019xpr, Liao:2017uzy, Dent:2017mpr, Dent:2019ueq, Miranda:2020syh, Suliga:2020jfa,Dutta:2020che, Dutta:2019nbn}. Since CEvNS cross section is orders of magnitude higher at low energies, the BSM searches can be done with relatively small detectors instead of the typical large neutrino detectors. \\ \\
Any significant deviations between the SM theory predictions and the experiment event rate of Eq.~({\ref{rate}) will indicate the presence of new physics and can be expressed as some conventional definition for the low-energy property. The deviation from the SM predictions can either be reflected as a change in the total event rate or a change in the shape of the recoil spectrum. For a given stopped pion production facility, experiments have control over choosing the baseline (the distance from the source to the detector), the angular placement of the detector with respect to the beam axis, and the nuclear target employed in the detector. These can be exploited to increase the sensitivity of a particular BSM scenario. An additional advantage of the stopped pion source is that one could exploit the timing structure of the neutrino background, see Fig.~\ref{fig:flux_2}, for a given BSM signal. 


\subsection{Non-standard Interactions of Neutrinos}

\begin{figure}
\centering
\includegraphics[width=0.5\columnwidth]{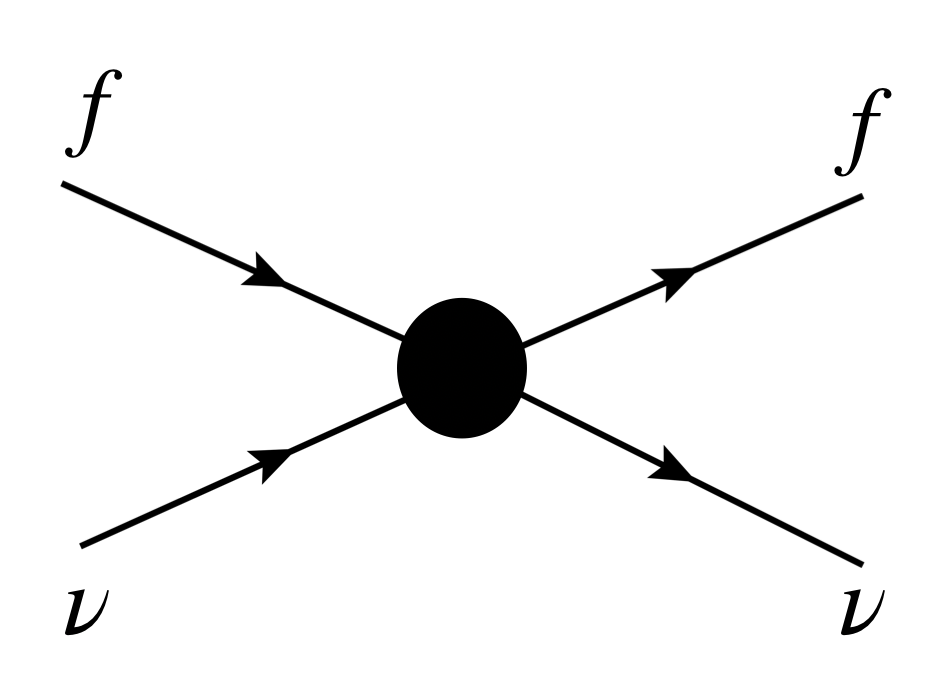}
\caption{A diagrammatic illustration of the neutral current non-standard neutrino interactions process where $f$ refers to SM fermions.}
\label{Feynman_Diagram_NSI}
\end{figure}
In the context of neutrino physics, the term Non-Standard Interactions (NSI) usually refers to the inclusion of four-fermion operators leading to modifications of the Wilson coefficients already present in the SM~\cite{Farzan:2017xzy, Wolfenstein:1977ue,  Proceedings:2019qno}. NSIs appear in several appealing SM extensions and provide a general effective field theory (EFT) framework to parameterize new physics in the neutrino sector. NSIs can be charged current or neutral current interactions with matter particles e, u, and/or d. Charged-current NSI leads to modification of both the production and detection of neutrinos but also leads to charge-lepton flavor violation while NC NSI does not modify production or detection when charged leptons are involved. \\ \\
In a model-independent approach, a useful parametrization of the possible effects of new physics at low energies is through the addition of higher-dimensional operators to the SM Lagrangian that respect the SM gauge group. The allowed set of operators includes four-fermion operators affecting neutrino production, propagation, and detection processes. For example, operators of the form

\begin{equation}
\mathcal{L}_{NSI}^{CC} ~=~ 2\sqrt{2}G_F~ \varepsilon_{\alpha\beta}^{ff', P}~\bar{\nu}_\alpha\gamma^\mu(1-\gamma_5)l_\beta~\bar{f'}\gamma_\mu f
\end{equation}
would induce non-standard charged-current (CC) production and detection processes for neutrinos of flavor $\alpha$, while operators such as

\begin{equation}
\mathcal{L}_{NSI}^{NC} ~=~ -2\sqrt{2}G_F~ \varepsilon_{\alpha\beta}^{fP}~\bar{\nu}_\alpha\gamma^\mu(1-\gamma_5)\nu_\beta~\bar{f}\gamma_\mu f
\end{equation}
would lead to flavor-changing neutral-current (NC) interactions of neutrinos with other fermions (if $\alpha = \beta$), or to a modified NC interaction rate with respect to the SM expectation (if $\alpha \ne \beta$). Here, $f$ and $f'$ refer to SM fermions. The parameters $\varepsilon$ describe the size of non-standard interactions relative to standard charged or neutral current weak interactions. \\ \\
In the SM, the weak charge of a nucleus only depends on the SM vector couplings to protons and neutrons, Eq.~(\ref{eq:weakcharge}), and is independent of the neutrino flavor. In the presence of NC NSI, this effective charge gets modified by the new operators introduced as

\begin{eqnarray}
Q^2_{W,NSI} & = & [(g_n^V+2\varepsilon_{\alpha\alpha}^{uV}+\varepsilon_{\alpha\alpha}^{dV}) N +(g_p^V+\varepsilon_{\alpha\alpha}^{uV}+2\varepsilon_{\alpha\alpha}^{dV})Z]^2 \nonumber \\
                      &    & + \sum_{\beta\neq\alpha}[(\varepsilon_{\alpha\beta}^{uV}+2\varepsilon_{\alpha\beta}^{dV}) N+(2\varepsilon_{\alpha\beta}^{uV}+\varepsilon_{\alpha\beta}^{dV})Z]^2
\end{eqnarray}
\\
Any deviation from SM predicted rates (plus the form factor uncertainty) signals a probing non-standard interaction of neutrinos. Vector couplings are characterized by the spin-independent combination while axial couplings are characterized by the orthogonal spin-dependent combination. Since typically in the CEvNS process, the axial contribution is negligible in comparison to the vector contribution (due to spin cancellation), we assume that spin-dependent axial NSI contributions are small. The effect of non-zero values of epsilons, which can be either positive or negative, can be either an enhancement or suppression of the CEvNS rate. Of course, some combinations of NSI parameter values for a given $N$ and $Z$ can result in the SM CEvNS rate. Therefore, a combination of CEvNS measurements on targets with different $N$ and $Z$ values can break any such accidental degeneracies (as well as cancel flux-related uncertainties). A stopped-pion source contains both electron and muon flavor; hence a CEvNS measurement gives direct access to all coupling coefficients except $\varepsilon_{\tau\tau}$. Furthermore, because at the SNS neutrino flavors can be separated by timing, see Fig.~\ref{fig:flux_2}, one can in principle probe electron and muon NSIs separately~\cite{Coloma:2017ncl, Liao:2017uzy, Dent:2017mpr}. \\ \\
The flavor-changing NSI in neutrino oscillation experiments leads to the appearance of new degeneracies involving standard oscillation parameters and NSI operators. These can affect DUNE's capability in extracting the CP-violation phase. Data from the CEvNS experiment to the global fits from oscillation breaks the degeneracies involving flavor-diagonal NSI, since CEvNS experiments can directly measure the neutrino-nucleus interactions for both electron and muon neutrinos. Thus constraining NSI parameters with CEvNS experiments can significantly improve the extraction of the CP-violation phase in DUNE~\cite{Coloma:2016gei, Coloma:2017egw}. 


\subsection{Neutrino Electromagnetic Properties}

In the SM, neutrino electromagnetic interactions are negligible. BSM processes can induce significant electromagnetic effects, these can be probed in CEvNS experiments. Since neutrinos oscillate and therefore have mass, it provides the best motivation for the existence of non-trivial neutrino electromagnetic properties such as a neutrino magnetic moment or a neutrino charge radius. \\ \\
The SM, minimally extended to allow massive neutrinos, predicts a very small magnetic moment of neutrino~\cite{ParticleDataGroup:2020ssz}

\begin{equation}
\mu_\nu = 3.2\times 10^{-19} \mu_B \left(\frac{m_{\nu}}{{\rm eV}}\right).
\end{equation}
\\
Within the minimum neutrino mass range of $10^{-4}$ to $1$ eV, the SM electron neutrino magnetic moment ranges [$3 \times 10^{-21}, 3 \times 10^{-19}$] $\mu_B$ for Dirac and  [$5 \times 10^{-25}, 8 \times 10^{-23}$] $\mu_B$ for Majorana case~\cite{Balantekin:2013sda}. The differential cross section in the presence of a neutrino magnetic moment adds incoherently to the Standard Model cross section due to the required spin-flip

\begin{equation}
\left( \frac{d \sigma}{dT}\right)_{\mathrm{tot}} = \left( \frac{d \sigma}{dT} \right)_{\mathrm{SM}} 
+ \left( \frac{d \sigma}{dT} 
\right)_{\mathrm{EM}} 
\end{equation}
\\
where the EM contribution has a characteristic $1/T$ dependence, while its strength is controlled by the size of the neutrino magnetic moment~\cite{Vogel:1989iv}

\begin{equation}
\left( \frac{d \sigma}{dT} \right)_{\mathrm{EM}}=\frac{\pi \alpha^2 \mu_{\nu}^{2}\,Z^{2}}{m_{e}^{2}}\left(\frac{1-T/E_{\nu}}{T}+\frac{T}{4E_\nu^2}\right) F_\text{ch}^{2}(q^{2})\, \\
\label{NMM-cross section}
\end{equation}
\\
where $m_e$ is the mass of the electron, $\alpha$ is the fine structure constant. $F_\text{ch}$, normalized as $F_\text{ch}(0)=1$, is the charge form factor of the nucleus that is known with high precision for many nuclei. In the presence of neutrino magnetic moments, the neutrino interaction cross section will be modified. The low-energy cross section is more sensitive to these small changes. Any measurement of magnetic moment larger than this would be a signature of BSM physics. Larger neutrino magnetic moments would also hint that neutrinos are Majorana particles~\cite{Bell:2006wi}. The current best limits on neutrino magnetic moment have been set by solar and astrophysical neutrinos~\cite{ParticleDataGroup:2020ssz, Borexino:2017fbd}. The strongest magnetic moment limits on $\nu_\mu$ are from LSND experiment~\cite{LSND:2001akn, Kosmas:2015sqa}. The experimental signature would be an enhancement of the rate at low recoil energy, both for scattering on electrons and on nuclei~\cite{Vogel:1989iv, Kosmas:2015sqa, Dodd:1991ni}. Therefore, to measure neutrino magnetic moment, the detector need to have a low energy threshold and good energy resolution.\\ \\
The impact of the neutrino charge radius, being a helicity-preserving quantity, is taken as a shift on the weak mixing angle according to 

\begin{equation}
\sin^2 \theta_W \rightarrow \sin^2 \theta_W + \frac{\sqrt{2} \pi \alpha}{3 G_F} \langle r_{\nu_\alpha}^2\rangle \, .
\label{eq:rv}
\end{equation}
\\
The neutrino charge radius has a small flavor-dependent effect on the CEvNS cross section. These can be measured in CEvNS experiments at stopped pion sources where muon- and electron-neutrino's CEvNS event rates can be separated using the timing structure, see Fig.~\ref{fig:flux_2}. The effects is expected to be of percent level~\cite{Papavassiliou:2005cs, Cadeddu:2018dux}. 


\subsection{Sterile Neutrino Oscillations}

One possible extension of the SM is the existence of sterile neutrinos, new neutrino states that do not interact with the SM weak interactions. These new gauge singlet fermions can be included as a minimal extension of the SM. There are several experimental anomalies, called short-baseline neutrino anomalies, that hint at the existence of sterile neutrinos. There are a number of experiments underway testing short-baseline anomalies. CEvNS is also an excellent tool for the search for sterile neutrino oscillations by setting up multiple identical detectors at different baselines from the neutrino production point. Flavor-blind neutral-currents can be used to probe the disappearance of active neutrinos. The signal would be both a distortion of recoil spectra and an overall rate suppression. \\ \\
Oscillation probabilities are simplest for monoenergetic sources and decay at rest provides a high-intensity monoenergetic source of $\nu_\mu$s, thus being a natural candidate for carrying out high statistics searches for $\nu_\mu$ disappearance. In this context, CEvNS has been recognized as being advantageous due to its relatively large cross section \cite{Anderson:2012pn, Formaggio:2011jt, Kosmas:2017zbh}. Working at lower energies allows for shorter baselines with equivalent L/E and consequently higher fluxes as compared to e.g., the decay in-flight experiments. In particular, the sensitivity to sterile neutrinos is maximized when deploying multiple detectors at different distance baselines in the range $\sim 20$--$40$ m~\cite{Anderson:2012pn}. This configuration can probe parameter space that is consistent with the $\sim$ eV mass-scale hinted at by LSND and MiniBooNE, thereby providing an independent test of sterile neutrino parameter space~\cite{Anderson:2012pn,Blanco:2019vyp}. 


\subsection{Accelerator Produced Light Dark Matter}

\begin{figure}
\centering
\includegraphics[width=0.6\columnwidth]{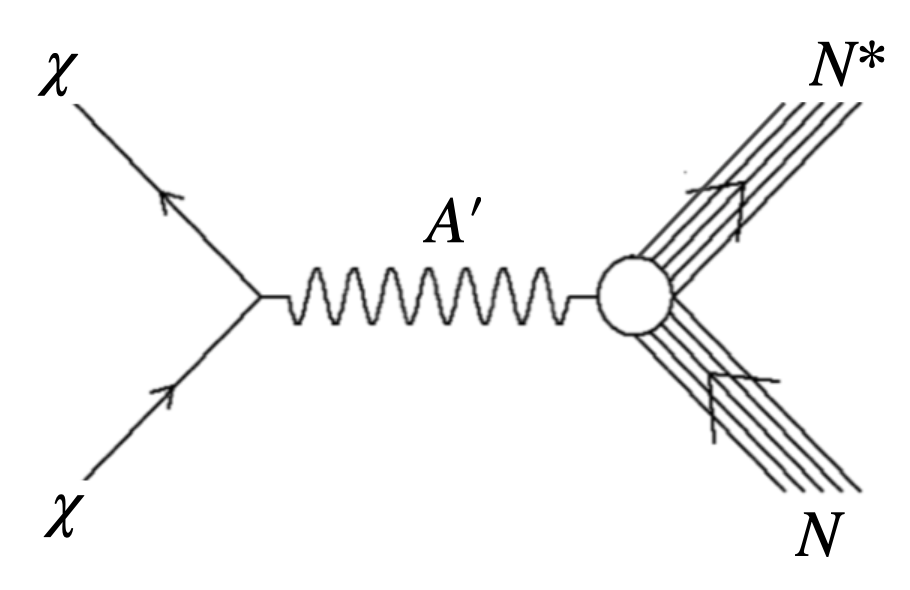}
\caption{A diagrammatic representation of the Dark Matter particle, $\chi$, scattering off a nucleus, $N$, mediated via a dark photon $A'$.}
\label{fig:DM_scattering}
\end{figure}

The stopped-pion facilities produce a copious amount of neutral and charged mesons as well as photons. These can decay into dark sector particles that can either scatter off or decay in the detector material via kinetic mixing with the SM particles~\cite{Dutta:2019nbn, deNiverville:2015mwa, deNiverville:2016rqh}. Dark matter candidate particle masses can be probed in the few to few hundred MeV range at these facilities. For example, within a dark photon model, the dark photon $A'$ undergoes kinetic mixing with the SM photon and can be described by the Lagrangian:
\begin{equation}
    \mathcal{L} \supset g_D A'_\mu \bar \chi \gamma^\mu \chi + e \epsilon Q_q A'_\mu \bar q \gamma^\mu q
\end{equation}
where $g_D$ is the dark coupling constant, $\epsilon$ is the mixing parameter, $Q_q$ is quark's electric charge. The dark photon can be produced at the stopped pion facilities via the processes of pion capture, pion decay, and the photons emerging from the cascades:

\begin{eqnarray}
    \pi^- + p &\rightarrow& n + A' \\
    \pi^+ + n &\rightarrow& p + A' \\
    \pi^0 &\rightarrow& \gamma + A'
\end{eqnarray}
The dark photons then decay into dark matter candidates: $A' \rightarrow \chi\bar{\chi}$. The signature in the detector is either elastic scattering with specific keV energy nuclear recoil signatures or inelastic scattering with specific MeV energy nuclear deexcitation gammas~\cite{Dutta:2022tav, Dutta:2023fij}. The process is schematically shown in Fig.~\ref{fig:DM_scattering}. \\ \\
The time structure of the beam is especially valuable for separating SM neutrino scattering recoils from the DM-induced recoils. Since the DM signal is expected to be prompt, the delayed neutrino signal can provide a powerful constraint on the neutrino background in the prompt window. The DM signal will also have a distinctive dependence on the direction with respect to the source, whereas CEvNS signals will be isotropic. For dark sector searches, event rate scaling, recoil spectrum shape, timing, and direction with respect to source are all helpful. \\ \\
Recent results from the COHERENT and Coherent CAPTAIN-Mills experiments have demonstrated how detectors capable of measuring coherent elastic neutrino-nucleus scattering (CEvNS) can also be used to set limits on vector portal and leptophobic DM at proton beam dumps. They also provide excellent opportunities to search for axion-like particles (ALPs)~\cite{Dent:2019ueq}.


\section{Summary}\label{sec:summary}

Neutrinos continue to provide a testing ground for the structure of the standard model and hints toward the physics beyond the standard model. Neutrinos of energies spanning over several orders of magnitude have been detected via various mechanisms ranging from inverse beta decay to scattering off quarks, nucleons, and nuclei. At MeV scales, there has been one elusive process, until a few years ago, known as coherent elastic neutrino-nucleus scattering that was theoretically predicted for over five decades ago but was never observed experimentally. The recent experimental observation of CEvNS by the COHERENT collaboration at a stopped pion neutrino source has inspired physicists across many subfields. This new way of detecting neutrinos has wider implications for border communities that span nuclear physics, particle physics, astrophysics, and beyond. Leveraging orders of magnitude higher CEvNS cross section, new physics can be searched with relatively small detectors. CEvNS, being a low-energy process, provides a natural window to study light, weakly-coupled, new physics in the neutrino sector. \\ \\
Neutrinos from stopped pions sources cover energies in the tens of MeVs and are almost optimal for studying CEvNS, finding a sweet spot where the CEvNS rate is high enough and recoil energies are more easily detectable above the threshold. So far, CEvNS is observed only at the decay at rest sources. In addition, the pulsed time structure
of the beam source provides a strong handle for suppressing the background for new physics searches. Several worldwide experimental programs have been or are being set up to detect CEvNS and BSM signals in the near future at stopped–pion neutrino sources (as well as with reactor sources where the CEvNS process is yet to be detected) with complementary detection technologies and physics goals, making it an emerging exciting avenue.\\ \\
Disentangling new physics signals in these experiments requires a precise understanding of the CEvNS SM scattering rate. At tree level, the theoretical uncertainty on the CEvNS cross section is driven by the uncertainty on the weak form factor of the nucleus. The charge density of a nucleus is strongly dominated by protons and has been extensively studied with impressive precision in elastic electron scattering experiments. While the neutron density, which CEvNS is most sensitive to, leads the overall uncertainties on the CEvNS rate. For non-zero spin nuclei, the axial-vector part adds an additional contribution that is often not included in CEvNS estimation. The CEvNS process also receives a few percent of radiative corrections, from electrons and muons running in loops introducing a non-trivial dependence on the momentum transfer due to their relatively light masses. Parity-violating electron scattering experiments offer complementary input and  provide a precise approach to experimentally probing weak form factors and neutron distribution. Although, the choice of the nuclear targets, so far in PVES and CEvNS experiments is not the same, since they are both driven by different physics motivations and have varied technical needs. \\ \\
CEvNS experiments at stopped pion sources are also sensitive to tens of MeV inelastic CC and NC neutrino-nucleus scattering processes. These processes have implications for supernova detection in future neutrino experiments. The interaction cross sections for these processes lack the $N^2$ enhancement associated with CEvNS and, therefore, tend to be at least one order of magnitude smaller than that of the CEvNS rate. The detectable final-state particles of these inelastic scattering have typical energies of the same order as the incident neutrino energies. The experimental requirements for CEvNS and inelastic signals are quite different, larger masses are needed for inelastic, as well as the dynamic range to record MeV-scale energy depositions, while very low thresholds are not required. DUNE’s capabilities of supernova neutrino detection
in the relevant tens-of-MeV neutrino energy range as well as the physics to be learned from a DUNE supernova burst detection will be limited by the lack of knowledge of the inelastic neutrino-argon cross section. The well-understood stopped-pion neutrino spectrum is a near-ideal tens of MeV neutrino source which provides a unique opportunity to study inelastic neutrino-nucleus cross sections at tens of MeVs.\\ \\
Since the uncertainty on the SM predicted CEvNS cross sections is relatively small, CEvNS cross section measurement allows testing of SM weak physics and in probing new physics signals. For a given stopped pion production facility, experiments can in principle choose the baseline, the angular placement
of the detector with respect to the beam axis, and the nuclear target employed in the detector to optimize the sensitivity of the primary physics goal of the experiment. An additional advantage of the stopped pion source is that one could exploit both timing and energy data. Any deviation from the SM predicted event rate either with a change in the total event rate or with a change in the shape of the recoil spectrum, could indicate new contributions to the interaction cross-section. In particular, the weak nuclear form factor of a nucleus and weak mixing angle can be probed. CEvNS, being a low-energy process, provides a natural window to study light, weakly-coupled, beyond the standard model physics in the neutrino sector. Several extensions of the SM can be explored at low energy. Since CEvNS cross section is orders of magnitude higher at low energies, the BSM searches can be done with the relatively small detector. In particular, NSIs, neutrino electromagnetic properties in terms of neutrino magnetic moment and neutrino charge radius, and sterile neutrinos can be studied. Stopped pion facilities are also a copious source of neutral and changed mesons as well as photons that allows proving several dark sector physics scenarios such as vector portal models, leptophobic dark matter as well as axion-like particle searches.


\section*{Acknowledgements}
V.P. thanks Oleksandr Tomalak, Pedro Machado and Ryan Plestid for discussions on Ref.~\cite{Tomalak:2020zfh}; Nils Van Dessel, Heather Ray and Natalie Jachowicz for discussion on Ref.~\cite{VanDessel:2020epd}; Bhaskar Dutta, Wei-Chih Huang and Jayden Newstead for discussion on Ref.~\cite{Dutta:2022tav}; colleagues from the Coherent CAPTAIN-Mills experiment for various discussions on the experimental scope of the CEvNS experiments - all of which have motivated the content of this review. This manuscript has been authored by Fermi Research Alliance, LLC under Contract No. DE-AC02-07CH11359 with the U.S. Department of Energy, Office of Science, Office of High Energy Physics.




\begin{thebibliography}{99}
\bibitem{Huber:2022lpm} P.~Huber, K.~Scholberg, E.~Worcester, J.~Asaadi, A.~B.~Balantekin, N.~Bowden, P.~Coloma, P.~B.~Denton, A.~de Gouv\^ea and L.~Fields, \textit{et al.} ``Snowmass Neutrino Frontier Report,'' arXiv:2211.08641 [hep-ex].
\bibitem{Balantekin:2022jrq} A.~B.~Balantekin, S.~Gardiner, K.~Mahn, T.~Mohayai, J.~Newby, V.~Pandey, J.~Zettlemoyer, J.~Asaadi, M.~Betancourt and D.~A.~Harris, \textit{et al.}
 ``Snowmass Neutrino Frontier: Neutrino Interaction Cross Sections (NF06) Topical Group Report,'' arXiv:2209.06872 [hep-ex].
\bibitem{deGouvea:2022gut} A.~de Gouv\^ea, I.~Mocioiu, S.~Pastore, L.~E.~Strigari, L.~Alvarez-Ruso, A.~M.~Ankowski, A.~B.~Balantekin, V.~Brdar, M.~Cadeddu and S.~Carey, \textit{et al.} ``Theory of Neutrino Physics -- Snowmass TF11 (aka NF08) Topical Group Report,'' arXiv:2209.07983 [hep-ph].
\bibitem{Acharya:2023swl} B.~Acharya, C.~Adams, A.~A.~Aleksandrova, K.~Alfonso, P.~An, S.~Bae\ss{}ler, A.~B.~Balantekin, P.~S.~Barbeau, F.~Bellini and V.~Bellini, \textit{et al.} ``Fundamental Symmetries, Neutrons, and Neutrinos (FSNN): Whitepaper for the 2023 NSAC Long Range Plan,'' arXiv:2304.03451 [nucl-ex].
\bibitem{Davis:1968cp} R.~Davis, Jr., D.~S.~Harmer and K.~C.~Hoffman, ``Search for neutrinos from the sun,'' Phys. Rev. Lett. \textbf{20}, 1205-1209 (1968).
\bibitem{Kamiokande-II:1987idp} K.~Hirata \textit{et al.} [Kamiokande-II], ``Observation of a Neutrino Burst from the Supernova SN 1987a,'' Phys. Rev. Lett. \textbf{58}, 1490-1493 (1987).
\bibitem{Bionta:1987qt} R.~M.~Bionta, G.~Blewitt, C.~B.~Bratton, D.~Casper, A.~Ciocio, R.~Claus, B.~Cortez, M.~Crouch, S.~T.~Dye and S.~Errede, \textit{et al.} ``Observation of a Neutrino Burst in Coincidence with Supernova SN 1987a in the Large Magellanic Cloud,'' Phys. Rev. Lett. \textbf{58}, 1494 (1987).
\bibitem{IceCube:2018cha} M.~G.~Aartsen \textit{et al.} [IceCube], ``Neutrino emission from the direction of the blazar TXS 0506+056 prior to the IceCube-170922A alert,'' Science \textbf{361}, no.6398, 147-151 (2018). 
\bibitem{Formaggio:2012cpf} J.~A.~Formaggio and G.~P.~Zeller, ``From eV to EeV: Neutrino Cross Sections Across Energy Scales,'' Rev. Mod. Phys. \textbf{84}, 1307-1341 (2012).
\bibitem{Stodolsky:1966zz} L.~Stodolsky, ``Application of Nuclear Coherence Properties to Elementary-Particle Reactions,'' Phys. Rev. \textbf{144}, 1145-1153 (1966).
\bibitem{Freedman:1973yd} D.~Z.~Freedman, ``Coherent Neutrino Nucleus Scattering as a Probe of the Weak Neutral Current,'' Phys. Rev. D \textbf{9}, 1389-1392 (1974).
\bibitem{Kopeliovich:1974mv} V.~B.~Kopeliovich and L.~L.~Frankfurt, ``Isotopic and chiral structure of neutral current,'' JETP Lett. \textbf{19}, 145-147 (1974).
\bibitem{COHERENT:2017ipa} D.~Akimov \textit{et al.} [COHERENT], ``Observation of Coherent Elastic Neutrino-Nucleus Scattering,'' Science \textbf{357}, no.6356, 1123-1126 (2017).
\bibitem{COHERENT:2018imc} D.~Akimov \textit{et al.} [COHERENT], ``COHERENT Collaboration data release from the first observation of coherent elastic neutrino-nucleus scattering,'' arXiv:1804.09459 [nucl-ex].
\bibitem{COHERENT:2019iyj} D.~Akimov \textit{et al.} [COHERENT], ``First Constraint on Coherent Elastic Neutrino-Nucleus Scattering in Argon,'' Phys. Rev. D \textbf{100}, no.11, 115020 (2019).
\bibitem{COHERENT:2020iec} D.~Akimov \textit{et al.} [COHERENT], ``First Measurement of Coherent Elastic Neutrino-Nucleus Scattering on Argon,'' Phys. Rev. Lett. \textbf{126}, no.1, 012002 (2021).
\bibitem{COHERENT:2021xmm} D.~Akimov \textit{et al.} [COHERENT], ``Measurement of the Coherent Elastic Neutrino-Nucleus Scattering Cross Section on CsI by COHERENT,'' Phys. Rev. Lett. \textbf{129}, no.8, 081801 (2022).
\bibitem{Barranco:2005yy} J.~Barranco, O.~G.~Miranda and T.~I.~Rashba, ``Probing new physics with coherent neutrino scattering off nuclei,'' JHEP \textbf{12}, 021 (2005).
\bibitem{Scholberg:2005qs} K.~Scholberg, ``Prospects for measuring coherent neutrino-nucleus elastic scattering at a stopped-pion neutrino source,'' Phys. Rev. D \textbf{73}, 033005 (2006).
\bibitem{Barranco:2007tz} J.~Barranco, O.~G.~Miranda and T.~I.~Rashba, ``Low energy neutrino experiments sensitivity to physics beyond the Standard Model,''
Phys. Rev. D \textbf{76}, 073008 (2007).
\bibitem{Dutta:2015nlo} B.~Dutta, Y.~Gao, R.~Mahapatra, N.~Mirabolfathi, L.~E.~Strigari and J.~W.~Walker, ``Sensitivity to oscillation with a sterile fourth generation neutrino from ultra-low threshold neutrino-nucleus coherent scattering,'' Phys. Rev. D \textbf{94}, no.9, 093002 (2016).
\bibitem{Lindner:2016wff} M.~Lindner, W.~Rodejohann and X.~J.~Xu, ``Coherent Neutrino-Nucleus Scattering and new Neutrino Interactions,'' JHEP \textbf{03}, 097 (2017).
\bibitem{Coloma:2017ncl} P.~Coloma, M.~C.~Gonzalez-Garcia, M.~Maltoni and T.~Schwetz, ``COHERENT Enlightenment of the Neutrino Dark Side,'' Phys. Rev. D \textbf{96}, no.11, 115007 (2017).
\bibitem{Farzan:2017xzy} Y.~Farzan and M.~Tortola, ``Neutrino oscillations and Non-Standard Interactions,'' Front. in Phys. \textbf{6}, 10 (2018).
\bibitem{Billard:2018jnl} J.~Billard, J.~Johnston and B.~J.~Kavanagh, ``Prospects for exploring New Physics in Coherent Elastic Neutrino-Nucleus Scattering,'' JCAP \textbf{11}, 016 (2018).
\bibitem{AristizabalSierra:2018eqm} D.~Aristizabal Sierra, V.~De Romeri and N.~Rojas, ``COHERENT analysis of neutrino generalized interactions,'' Phys. Rev. D \textbf{98}, 075018 (2018).
\bibitem{Brdar:2018qqj} V.~Brdar, W.~Rodejohann and X.~J.~Xu, ``Producing a new Fermion in Coherent Elastic Neutrino-Nucleus Scattering: from Neutrino Mass to Dark Matter,'' JHEP \textbf{12}, 024 (2018).
\bibitem{Abdullah:2018ykz} M.~Abdullah, J.~B.~Dent, B.~Dutta, G.~L.~Kane, S.~Liao and L.~E.~Strigari, ``Coherent elastic neutrino nucleus scattering as a probe of a Z' through kinetic and mass mixing effects,'' Phys. Rev. D \textbf{98}, no.1, 015005 (2018).
\bibitem{AristizabalSierra:2019zmy} D.~Aristizabal Sierra, J.~Liao and D.~Marfatia, ``Impact of form factor uncertainties on interpretations of coherent elastic neutrino-nucleus scattering data,'' JHEP \textbf{06}, 141 (2019).
\bibitem{Miranda:2019skf} O.~G.~Miranda, G.~Sanchez Garcia and O.~Sanders, ``Coherent elastic neutrino-nucleus scattering as a precision test for the Standard Model and beyond: the COHERENT proposal case,'' Adv. High Energy Phys. \textbf{2019}, 3902819 (2019).
\bibitem{Bell:2019egg} N.~F.~Bell, J.~B.~Dent, J.~L.~Newstead, S.~Sabharwal and T.~J.~Weiler, ``Migdal effect and photon bremsstrahlung in effective field theories of dark matter direct detection and coherent elastic neutrino-nucleus scattering,'' Phys. Rev. D \textbf{101}, no.1, 015012 (2020).
\bibitem{AristizabalSierra:2019ufd} D.~Aristizabal Sierra, V.~De Romeri and N.~Rojas, ``CP violating effects in coherent elastic neutrino-nucleus scattering processes,''
JHEP \textbf{09}, 069 (2019).
\bibitem{Cadeddu:2019eta} M.~Cadeddu, F.~Dordei, C.~Giunti, Y.~F.~Li and Y.~Y.~Zhang, ``Neutrino, electroweak, and nuclear physics from COHERENT elastic neutrino-nucleus scattering with refined quenching factor,'' Phys. Rev. D \textbf{101}, no.3, 033004 (2020).
\bibitem{Coloma:2019mbs} P.~Coloma, I.~Esteban, M.~C.~Gonzalez-Garcia and M.~Maltoni, ``Improved global fit to Non-Standard neutrino Interactions using COHERENT energy and timing data,'' JHEP \textbf{02}, 023 (2020).
\bibitem{Canas:2019fjw} B.~C.~Canas, E.~A.~Garces, O.~G.~Miranda, A.~Parada and G.~Sanchez Garcia, ``Interplay between nonstandard and nuclear constraints in coherent elastic neutrino-nucleus scattering experiments,'' Phys. Rev. D \textbf{101}, no.3, 035012 (2020).
\bibitem{Dutta:2019eml} B.~Dutta, S.~Liao, S.~Sinha and L.~E.~Strigari, ``Searching for Beyond the Standard Model Physics with COHERENT Energy and Timing Data,''
Phys. Rev. Lett. \textbf{123}, no.6, 061801 (2019).
\bibitem{Denton:2020hop} P.~B.~Denton and J.~Gehrlein, ``A Statistical Analysis of the COHERENT Data and Applications to New Physics,'' JHEP \textbf{04}, 266 (2021).
\bibitem{Skiba:2020msb} W.~Skiba and Q.~Xia, ``Electroweak constraints from the COHERENT experiment,'' JHEP \textbf{10}, 102 (2022).
\bibitem{Cadeddu:2020nbr} M.~Cadeddu, N.~Cargioli, F.~Dordei, C.~Giunti, Y.~F.~Li, E.~Picciau and Y.~Y.~Zhang, ``Constraints on light vector mediators through coherent elastic neutrino nucleus scattering data from COHERENT,'' JHEP \textbf{01}, 116 (2021).
\bibitem{Canas:2018rng} B.~C.~Ca\~nas, E.~A.~Garc\'es, O.~G.~Miranda and A.~Parada, ``Future perspectives for a weak mixing angle measurement in coherent elastic neutrino nucleus scattering experiments,'' Phys. Lett. B \textbf{784}, 159-162 (2018).
\bibitem{Alonso:2010fs} J.~Alonso, F.~T.~Avignone, W.~A.~Barletta, R.~Barlow, H.~T.~Baumgartner, A.~Bernstein, E.~Blucher, L.~Bugel, L.~Calabretta and L.~Camilleri, \textit{et al.} ``Expression of Interest for a Novel Search for CP Violation in the Neutrino Sector: DAEdALUS,'' arXiv:1006.0260 [physics.ins-det].
\bibitem{Tomalak:2021lif} O.~Tomalak, ``Radiative (anti)neutrino energy spectra from muon, pion, and kaon decays,'' Phys. Lett. B \textbf{829}, 137108 (2022).
\bibitem{Ishikawa:2018rlv} T.~Ishikawa, N.~Nakazawa and Y.~Yasui, ``Numerical calculation of the full two-loop electroweak corrections to muon (g-2),'' Phys. Rev. D \textbf{99}, no.7, 073004 (2019).
\bibitem{Abdullah:2020iiv} M.~Abdullah, D.~Aristizabal Sierra, B.~Dutta and L.~E.~Strigari, ``Coherent Elastic Neutrino-Nucleus Scattering with directional detectors,'' Phys. Rev. D \textbf{102}, no.1, 015009 (2020).
\bibitem{Hofstadter:1956qs} R.~Hofstadter, ``Electron scattering and nuclear structure,'' Rev. Mod. Phys. \textbf{28}, 214-254 (1956).
\bibitem{DeVries:1987atn} H.~De Vries, C.~W.~De Jager and C.~De Vries, ``Nuclear charge and magnetization density distribution parameters from elastic electron scattering,'' Atom. Data Nucl. Data Tabl. \textbf{36}, 495-536 (1987).
\bibitem{Fricke:1995zz} G.~Fricke, C.~Bernhardt, K.~Heilig, L.~A.~Schaller, L.~Schellenberg, E.~B.~Shera and C.~W.~de Jager, ``Nuclear Ground State Charge Radii from Electromagnetic Interactions,'' Atom. Data Nucl. Data Tabl. \textbf{60}, 177-285 (1995).
\bibitem{Angeli:2013epw} I.~Angeli and K.~P.~Marinova, ``Table of experimental nuclear ground state charge radii: An update,'' Atom. Data Nucl. Data Tabl. \textbf{99}, no.1, 69-95 (2013).
\bibitem{Thiel:2019tkm} M.~Thiel, C.~Sfienti, J.~Piekarewicz, C.~J.~Horowitz and M.~Vanderhaeghen, ``Neutron skins of atomic nuclei: per aspera ad astra,'' J. Phys. G \textbf{46}, no.9, 093003 (2019).
\bibitem{Donnelly:1989qs} T.~W.~Donnelly, J.~Dubach and I.~Sick, ``Isospin Dependences in Parity Violating Electron Scattering,'' Nucl. Phys. A \textbf{503}, 589-631 (1989).
\bibitem{Helm:1956} R.~H.~Helm, ``Inelastic and Elastic Scattering of 187-Mev Electrons from Selected Even-Even Nuclei,'' Phys. Rev. \textbf{104}, 1466-1475 (1956).
\bibitem{KN:1999} S.~Klein and J.~Nystrand, ``Exclusive vector meson production in relativistic heavy ion collisions,'' Phys. Rev. C \textbf{60}, 014903 (1999).
\bibitem{Duda:2006uk} G.~Duda, A.~Kemper and P.~Gondolo, ``Model Independent Form Factors for Spin Independent Neutralino-Nucleon Scattering from Elastic Electron Scattering Data,'' JCAP \textbf{04}, 012 (2007).
\bibitem{Lewin:1995rx} J.~D.~Lewin and P.~F.~Smith, ``Review of mathematics, numerical factors, and corrections for dark matter experiments based on elastic nuclear recoil,''
Astropart. Phys. \textbf{6}, 87-112 (1996).
\bibitem{Papoulias:2019xaw} D.~K.~Papoulias, T.~S.~Kosmas and Y.~Kuno, ``Recent probes of standard and non-standard neutrino physics with nuclei,'' Front. in Phys. \textbf{7}, 191 (2019).
\bibitem{Payne:2019wvy} C.~G.~Payne, S.~Bacca, G.~Hagen, W.~Jiang and T.~Papenbrock, ``Coherent elastic neutrino-nucleus scattering on $^{40}$Ar from first principles,'' Phys. Rev. C \textbf{100}, no.6, 061304 (2019).
\bibitem{Hoferichter:2020osn} M.~Hoferichter, J.~Men\'endez and A.~Schwenk, ``Coherent elastic neutrino-nucleus scattering: EFT analysis and nuclear responses,'' Phys. Rev. D \textbf{102}, no.7, 074018 (2020).
\bibitem{Yang:2019pbx} J.~Yang, J.~A.~Hernandez and J.~Piekarewicz, ``Electroweak probes of ground state densities,'' Phys. Rev. C \textbf{100}, no.5, 054301 (2019).
\bibitem{VanDessel:2020epd} N.~Van Dessel, V.~Pandey, H.~Ray and N.~Jachowicz, ``Cross sections for coherent elastic and inelastic neutrino-nucleus scattering,'' Universe \textbf{9}, 207 (2023).
\bibitem{Tomalak:2020zfh} O.~Tomalak, P.~Machado, V.~Pandey and R.~Plestid, ``Flavor-dependent radiative corrections in coherent elastic neutrino-nucleus scattering,'' JHEP \textbf{02}, 097 (2021).
\bibitem{Horowitz:1999fk} C.~J.~Horowitz, S.~J.~Pollock, P.~A.~Souder and R.~Michaels, ``Parity violating measurements of neutron densities,''
Phys. Rev. C \textbf{63}, 025501 (2001).
\bibitem{Yennie:1954zz} D.~R.~Yennie, D.~G.~Ravenhall and R.~N.~Wilson, ``Phase-Shift Calculation of High-Energy Electron Scattering,'' Phys. Rev. \textbf{95}, 500-512 (1954).
\bibitem{Yennie:1965zz} D.~R.~Yennie, F.~L.~Boos and D.~G.~Ravenhall, ``Analytic Distorted-Wave Approximation for High-Energy Electron Scattering Calculations,''
Phys. Rev. \textbf{137}, B882-B903 (1965).
\bibitem{Kim:1996ua} K.~S.~Kim, L.~E.~Wright, Y.~Jin and D.~W.~Kosik, ``Approximate treatment of electron Coulomb distortion in quasielastic (e,e') reactions,''
Phys. Rev. C \textbf{54}, 2515 (1996).
\bibitem{Kim:2001sq} K.~S.~Kim, L.~E.~Wright and D.~A.~Resler, ``Coulomb distortion effects for electron or positron induced (e, e-prime) reactions in the quasielastic region,'' Phys. Rev. C \textbf{64}, 044607 (2001).
\bibitem{Horowitz:1998vv} C.~J.~Horowitz, ``Parity violating elastic electron scattering and Coulomb distortions,'' Phys. Rev. C \textbf{57}, 3430-3436 (1998).
\bibitem{Abrahamyan:2012gp} S.~Abrahamyan, Z.~Ahmed, H.~Albataineh, K.~Aniol, D.~S.~Armstrong, W.~Armstrong, T.~Averett, B.~Babineau, A.~Barbieri and V.~Bellini \textit{et al.}, ``Measurement of the Neutron Radius of 208Pb Through Parity-Violation in Electron Scattering,'' Phys. Rev. Lett. \textbf{108}, 112502 (2012).
\bibitem{Horowitz:2012tj} C.~J.~Horowitz, Z.~Ahmed, C.~M.~Jen, A.~Rakhman, P.~A.~Souder, M.~M.~Dalton, N.~Liyanage, K.~D.~Paschke, K.~Saenboonruang and R.~Silwal, \textit{et al.}, ``Weak charge form factor and radius of 208Pb through parity violation in electron scattering,'' Phys. Rev. C \textbf{85}, 032501 (2012).
\bibitem{Kumar:2020ejz} K.~S.~Kumar [PREX and CREX], ``Electroweak probe of neutron skins of nuclei,'' Annals Phys. \textbf{412}, 168012 (2020).
\bibitem{Becker:2018ggl} D.~Becker, R.~Bucoveanu, C.~Grzesik, K.~Imai, R.~Kempf, K.~Imai, M.~Molitor, A.~Tyukin, M.~Zimmermann and D.~Armstrong \textit{et al.}, ``The P2 experiment,'' Eur. Phys. J. A \textbf{54}, no.11, 208 (2018).
\bibitem{Amanik:2009zz} P.~S.~Amanik and G.~C.~McLaughlin, ``Nuclear neutron form factor from neutrino nucleus coherent elastic scattering,'' J. Phys. G \textbf{36}, 015105 (2009).
\bibitem{Patton:2012jr} K.~Patton, J.~Engel, G.~C.~McLaughlin and N.~Schunck, ``Neutrino-nucleus coherent scattering as a probe of neutron density distributions,'' Phys. Rev. C \textbf{86}, 024612 (2012).
\bibitem{Cadeddu:2017etk} M.~Cadeddu, C.~Giunti, Y.~F.~Li and Y.~Y.~Zhang, ``Average CsI neutron density distribution from COHERENT data,'' Phys. Rev. Lett. \textbf{120}, no.7, 072501 (2018).
\bibitem{VanDessel:2019obk} N.~Van Dessel, A.~Nikolakopoulos and N.~Jachowicz,``Lepton kinematics in low energy neutrino-Argon interactions,'' Phys. Rev. C \textbf{101}, no.4, 045502 (2020).
\bibitem{Pandey:2014tza} V.~Pandey, N.~Jachowicz, T.~Van Cuyck, J.~Ryckebusch and M.~Martini, ``Low-energy excitations and quasielastic contribution to electron-nucleus and neutrino-nucleus scattering in the continuum random-phase approximation,'' Phys. Rev. C \textbf{92}, no.2, 024606 (2015).
\bibitem{OConnell:1972edu} J.~S.~O' Connell, T.~W.~Donnelly and J.~D.~Walecka, ``Semileptonic weak interactions with $C^{12}$,'' Phys. Rev. C \textbf{6}, 719-733 (1972).
\bibitem{Walecka:1995} J.~D.~Walecka, ``Theoretical Nuclear and Subnuclear Physics, ''  Oxford University Press, New York (1995).
\bibitem{Jachowicz:2002rr} N.~Jachowicz, K.~Heyde, J.~Ryckebusch and S.~Rombouts, ``Continuum random phase approximation approach to charged current neutrino nucleus scattering,''
Phys. Rev. C \textbf{65}, 025501 (2002).
\bibitem{Pandey:2016jju} V.~Pandey, N.~Jachowicz, M.~Martini, R.~Gonz\'alez-Jim\'enez, J.~Ryckebusch, T.~Van Cuyck and N.~Van Dessel, ``Impact of low-energy nuclear excitations on neutrino-nucleus scattering at MiniBooNE and T2K kinematics,'' Phys. Rev. C \textbf{94}, no.5, 054609 (2016).
\bibitem{VanDessel:2019atx} N.~Van Dessel, N.~Jachowicz and A.~Nikolakopoulos, ``Forbidden transitions in neutral and charged current interactions between low-energy neutrinos and Argon,'' Phys. Rev. C \textbf{100}, no.5, 055503 (2019).
\bibitem{DUNE:2020zfm} B.~Abi \textit{et al.} [DUNE], ``Supernova neutrino burst detection with the Deep Underground Neutrino Experiment,'' Eur. Phys. J. C \textbf{81}, no.5, 423 (2021).
\bibitem{DUNE:2023rtr} A.~Abed Abud \textit{et al.} [DUNE], ``Impact of cross-section uncertainties on supernova neutrino spectral parameter fitting in the Deep Underground Neutrino Experiment,'' arXiv:2303.17007 [hep-ex].
\bibitem{DUNE:2020lwj} B.~Abi \textit{et al.} [DUNE], ``Deep Underground Neutrino Experiment (DUNE), Far Detector Technical Design Report, Volume I Introduction to DUNE,'' JINST \textbf{15}, no.08, T08008 (2020).
\bibitem{DUNE:2020ypp} B.~Abi \textit{et al.} [DUNE], ``Deep Underground Neutrino Experiment (DUNE), Far Detector Technical Design Report, Volume II: DUNE Physics,'' [arXiv:2002.03005 [hep-ex]].
\bibitem{DUNE:2020mra} B.~Abi \textit{et al.} [DUNE], ``Deep Underground Neutrino Experiment (DUNE), Far Detector Technical Design Report, Volume III: DUNE Far Detector Technical Coordination,'' JINST \textbf{15}, no.08, T08009 (2020).
\bibitem{DUNE:2020txw} B.~Abi \textit{et al.} [DUNE], ``Deep Underground Neutrino Experiment (DUNE), Far Detector Technical Design Report, Volume IV: Far Detector Single-phase Technology,'' JINST \textbf{15}, no.08, T08010 (2020).
\bibitem{Scholberg:2012id} K.~Scholberg, ``Supernova Neutrino Detection,'' Ann. Rev. Nucl. Part. Sci. \textbf{62}, 81-103 (2012).
\bibitem{CCM} A.~A.~Aguilar-Arevalo \textit{et al.} [CCM], ``First dark matter search results from Coherent CAPTAIN-Mills,'' Phys. Rev. D \textbf{106}, no.1, 012001 (2022).
\bibitem{Willis:1980pj} S.~E.~Willis, V.~W.~Hughes, P.~Nemethy, R.~L.~Burman, D.~R.~Cochran, J.~S.~Frank, R.~P.~Redwine, J.~Duclos, H.~Kaspar and C.~K.~Hargrove \textit{et al.}, ``A Neutrino Experiment to Test the Nature of Muon Number Conservation,'' Phys. Rev. Lett. \textbf{44}, 522 (1980) [erratum: Phys. Rev. Lett. \textbf{44}, 903 (1980); erratum: Phys. Rev. Lett. \textbf{45}, 1370 (1980)]. 
\bibitem{KARMEN:1998xmo} B.~Armbruster \textit{et al.} [KARMEN], ``Measurement of the weak neutral current excitation C-12(nu(mu) nu'(mu))C*-12(1+,1,15.1-MeV) at E(nu(mu)) = 29.8-MeV,'' Phys. Lett. B \textbf{423}, 15-20 (1998).
\bibitem{KARMEN:1991vkr} G.~Drexlin \textit{et al.} [KARMEN], ``First observation of the neutral current nuclear excitation C-12 (nu, nu-prime) C-12* (1+, 1).,'' Phys. Lett. B \textbf{267}, 321-324 (1991).
\bibitem{Maschuw:1998qh} R.~Maschuw [KARMEN], ``Neutrino spectroscopy with KARMEN,'' Prog. Part. Nucl. Phys. \textbf{40}, 183-192 (1998).
\bibitem{Zeitnitz:1994kz} B.~Zeitnitz [KARMEN], ``KARMEN: Neutrino physics at ISIS,'' Prog. Part. Nucl. Phys. \textbf{32}, 351-373 (1994).
\bibitem{Krakauer:1991rf} D.~A.~Krakauer, R.~L.~Talaga, R.~C.~Allen, H.~H.~Chen, R.~Hausammann, W.~P.~Lee, H.~J.~Mahler, X.~Q.~Lu, K.~C.~Wang and T.~J.~Bowles \textit{et al.}, ``Experimental study of neutrino absorption on carbon,'' Phys. Rev. C \textbf{45}, 2450-2463 (1992).
\bibitem{LSND:2001fbw} L.~B.~Auerbach \textit{et al.} [LSND], ``Measurements of charged current reactions of nu(e) on 12-C,'' Phys. Rev. C \textbf{64}, 065501 (2001).
\bibitem{LSND:2002oco} L.~B.~Auerbach \textit{et al.} [LSND], ``Measurements of charged current reactions of muon neutrinos on C-12,'' Phys. Rev. C \textbf{66}, 015501 (2002).
\bibitem{Distel:2002ch} J.~R.~Distel, B.~T.~Cleveland, K.~Lande, C.~K.~Lee, P.~S.~Wildenhain, G.~E.~Allen and R.~L.~Burman, ``Measurement of the cross-section for the reaction I-127 (nu(e), e-) Xe-127(bound states) with neutrinos from the decay of stopped muons,'' Phys. Rev. C \textbf{68}, 054613 (2003).
\bibitem{COHERENT:2023ffx} P.~An \textit{et al.} [COHERENT], ``Measurement of the inclusive electron-neutrino charged-current cross section on ${}^{127}$I with the COHERENT NaI$\nu$E detector,'' arXiv:2305.19594 [nucl-ex].
\bibitem{COHERENT:2022eoh} P.~An \textit{et al.} [COHERENT], ``Measurement of ${}^{nat}$Pb($\nu_e$,X$n$) production with a stopped-pion neutrino source,''arXiv:2212.11295 [hep-ex].
\bibitem{CONNIE} A.~Aguilar-Arevalo \textit{et al.} [CONNIE], ``Results of the Engineering Run of the Coherent Neutrino Nucleus Interaction Experiment (CONNIE),'' JINST \textbf{11}, no.07, P07024 (2016).
\bibitem{MINER} G.~Agnolet \textit{et al.} [MINER], ``Background Studies for the MINER Coherent Neutrino Scattering Reactor Experiment,'' Nucl. Instrum. Meth. A \textbf{853}, 53-60 (2017).
\bibitem{vGEN} V.~Belov, V.~Brudanin, V.~Egorov, D.~Filosofov, M.~Fomina, Y.~Gurov, L.~Korotkova, A.~Lubashevskiy, D.~Medvedev and R.~Pritula \textit{et al.}, ``The \ensuremath{\nu}GeN experiment at the Kalinin Nuclear Power Plant,'' JINST \textbf{10}, no.12, P12011 (2015).
\bibitem{NUCLEUS} R.~Strauss, J.~Rothe, G.~Angloher, A.~Bento, A.~G\"utlein, D.~Hauff, H.~Kluck, M.~Mancuso, L.~Oberauer and F.~Petricca \textit{et al.}, ``The $\nu$-cleus experiment: A gram-scale fiducial-volume cryogenic detector for the first detection of coherent neutrino-nucleus scattering,'' Eur. Phys. J. C \textbf{77}, 506 (2017).
\bibitem{RICOCHET} J.~Billard, R.~Carr, J.~Dawson, E.~Figueroa-Feliciano, J.~A.~Formaggio, J.~Gascon, S.~T.~Heine, M.~De Jesus, J.~Johnston and T.~Lasserre \textit{et al.}, ``Coherent Neutrino Scattering with Low Temperature Bolometers at Chooz Reactor Complex,'' J. Phys. G \textbf{44}, no.10, 105101 (2017).
\bibitem{TEXONO} H.~T.~Wong, ``Neutrino-nucleus coherent scattering and dark matter searches with sub-keV germanium detector,'' Nucl. Phys. A \textbf{844}, 229C-233C (2010).
\bibitem{NEON} J.~J.~Choi \textit{et al.} [NEON], ``Exploring coherent elastic neutrino-nucleus scattering using reactor electron antineutrinos in the NEON experiment,'' Eur. Phys. J. C \textbf{83}, no.3, 226 (2023).
\bibitem{vIOLETA} C.~Awe \textit{et al.} [CHANDLER, CONNIE, CONUS, Daya Bay, JUNO, MTAS, NEOS, NuLat, PROSPECT, RENO, Ricochet, ROADSTR Near-Field Working Group, SoLid, Stereo, Valencia-Nantes TAGS, vIOLETA and WATCHMAN], ``High Energy Physics Opportunities Using Reactor Antineutrinos,'' arXiv:2203.07214 [hep-ex].
\bibitem{Barbeau:2021exu} P.~S.~Barbeau, Y.~Efremenko and K.~Scholberg, ``COHERENT at the Spallation Neutron Source,'' arXiv:2111.07033 [hep-ex].
\bibitem{Asaadi:2022ojm} J.~Asaadi, P.~S.~Barbeau, B.~Bodur, A.~Bross, E.~Conley, Y.~Efremenko, M.~Febbraro, A.~Galindo-Uribarri, S.~Gardiner and D.~Gonzalez-Diaz \textit{et al.}, ``Physics Opportunities in the ORNL Spallation Neutron Source Second Target Station Era,'' arXiv:2209.02883 [hep-ex].
\bibitem{Henderson:2014paa} S.~Henderson, ``The Spallation Neutron Source accelerator system design,'' Nucl. Instrum. Meth. A \textbf{763}, 610-673 (2014).
\bibitem{Haines:2014kna} J.~R.~Haines, T.~J.~McManamy, T.~A.~Gabriel, R.~E.~Battle, K.~K.~Chipley, J.~A.~Crabtree, L.~L.~Jacobs, D.~C.~Lousteau, M.~J.~Rennich and B.~W.~Riemer, ``Spallation neutron source target station design, development, and commissioning,'' Nucl. Instrum. Meth. A \textbf{764}, 94-115 (2014).
\bibitem{CCM:2021yzc} A.~A.~Aguilar-Arevalo \textit{et al.} [CCM], ``First Leptophobic Dark Matter Search from the Coherent\textendash{}CAPTAIN-Mills Liquid Argon Detector,'' Phys. Rev. Lett. \textbf{129}, no.2, 021801 (2022).
\bibitem{CCM:2021lhc} A.~A.~Aguilar-Arevalo \textit{et al.} [CCM], ``Axion-Like Particles at Coherent CAPTAIN-Mills,'' arXiv:2112.09979 [hep-ph].
\bibitem{Ajimura:2017fld} S.~Ajimura, M.~K.~Cheoun, J.~H.~Choi, H.~Furuta, M.~Harada, S.~Hasegawa, Y.~Hino, T.~Hiraiwa, E.~Iwai and S.~Iwata \textit{et al.}, ``Technical Design Report (TDR): Searching for a Sterile Neutrino at J-PARC MLF (E56, JSNS2),'' arXiv:1705.08629 [physics.ins-det].
\bibitem{Ajimura:2020qni} S.~Ajimura, M.~Botran, J.~H.~Choi, J.~W.~Choi, M.~K.~Cheoun, T.~Dodo, H.~Furuta, J.~Goh, K.~Haga and M.~Harada, \textit{et al.} ``Proposal: JSNS$^2$-II,'' arXiv:2012.10807 [hep-ex].
\bibitem{Baxter:2019mcx} D.~Baxter, J.~I.~Collar, P.~Coloma, C.~E.~Dahl, I.~Esteban, P.~Ferrario, J.~J.~Gomez-Cadenas, M.~C.~Gonzalez-Garcia, A.~R.~L.~Kavner and C.~M.~Lewis \textit{et al.}, ``Coherent Elastic Neutrino-Nucleus Scattering at the European Spallation Source,'' JHEP \textbf{02}, 123 (2020).
\bibitem{Toups:2022yxs} M.~Toups, R.~G.~Van de Water, B.~Batell, S.~J.~Brice, P.~deNiverville, B.~Dutta, J.~Eldred, T.~Hapitas, R.~Harnik and A.~Karthikeyan \textit{et al.}, ``PIP2-BD: GeV Proton Beam Dump at Fermilab's PIP-II Linac,'' arXiv:2203.08079 [hep-ex].
\bibitem{Huang:2019ene} X.~R.~Huang and L.~W.~Chen, ``Neutron Skin in CsI and Low-Energy Effective Weak Mixing Angle from COHERENT Data,''
Phys. Rev. D \textbf{100}, no.7, 071301 (2019).
\bibitem{Bernabeu:2002nw} J.~Bernabeu, J.~Papavassiliou and J.~Vidal, ``On the observability of the neutrino charge radius,'' Phys. Rev. Lett. \textbf{89}, 101802 (2002) [erratum: Phys. Rev. Lett. \textbf{89}, 229902 (2002)].
\bibitem{Bernabeu:2002pd} J.~Bernabeu, J.~Papavassiliou and J.~Vidal, ``The Neutrino charge radius is a physical observable,'' Nucl. Phys. B \textbf{680}, 450-478 (2004).
\bibitem{Papavassiliou:2005cs} J.~Papavassiliou, J.~Bernabeu and M.~Passera, ``Neutrino-nuclear coherent scattering and the effective neutrino charge radius,'' PoS \textbf{HEP2005}, 192 (2006).
\bibitem{Cadeddu:2018dux} M.~Cadeddu, C.~Giunti, K.~A.~Kouzakov, Y.~F.~Li, Y.~Y.~Zhang and A.~I.~Studenikin, ``Neutrino Charge Radii From Coherent Elastic Neutrino-nucleus Scattering,'' Phys. Rev. D \textbf{98}, no.11, 113010 (2018) [erratum: Phys. Rev. D \textbf{101}, no.5, 059902 (2020)].
\bibitem{Co:2020gwl} G.~Co', M.~Anguiano and A.~M.~Lallena, ``Nuclear structure uncertainties in coherent elastic neutrino-nucleus scattering,'' JCAP \textbf{04}, 044 (2020).
\bibitem{Ciuffoli:2018qem} E.~Ciuffoli, J.~Evslin, Q.~Fu and J.~Tang, ``Extracting nuclear form factors with coherent neutrino scattering,'' Phys. Rev. D \textbf{97}, no.11, 113003 (2018).
\bibitem{Papoulias:2019lfi} D.~K.~Papoulias, T.~S.~Kosmas, R.~Sahu, V.~K.~B.~Kota and M.~Hota, ``Constraining nuclear physics parameters with current and future COHERENT data,'' Phys. Lett. B \textbf{800}, 135133 (2020).
\bibitem{Fattoyev:2017jql} F.~J.~Fattoyev, J.~Piekarewicz and C.~J.~Horowitz, ``Neutron Skins and Neutron Stars in the Multimessenger Era,'' Phys. Rev. Lett. \textbf{120} (2018) no.17, 172702.
\bibitem{Reed:2021nqk} B.~T.~Reed, F.~J.~Fattoyev, C.~J.~Horowitz and J.~Piekarewicz, ``Implications of PREX-2 on the Equation of State of Neutron-Rich Matter,'' Phys. Rev. Lett. \textbf{126} (2021) no.17, 172503.
\bibitem{Lattimer:2012xj} J.~M.~Lattimer and Y.~Lim, ``Constraining the Symmetry Parameters of the Nuclear Interaction,'' Astrophys. J. \textbf{771} (2013), 51.
\bibitem{Hebeler:2013nza} K.~Hebeler, J.~M.~Lattimer, C.~J.~Pethick and A.~Schwenk, ``Equation of state and neutron star properties constrained by nuclear physics and observation,''
Astrophys. J. \textbf{773} (2013), 11.
\bibitem{Hagen:2015yea} G.~Hagen, A.~Ekstr\"om, C.~Forss\'en, G.~R.~Jansen, W.~Nazarewicz, T.~Papenbrock, K.~A.~Wendt, S.~Bacca, N.~Barnea and B.~Carlsson \textit{et al.}, ``Neutron and weak-charge distributions of the $^{48}$Ca nucleus,'' Nature Phys. \textbf{12} (2015) no.2, 186-190.
\bibitem{Abdullah:2022zue} M.~Abdullah, H.~Abele, D.~Akimov, G.~Angloher, D.~Aristizabal Sierra, C.~Augier, A.~B.~Balantekin, L.~Balogh, P.~S.~Barbeau and L.~Baudis \textit{et al.}, ``Coherent elastic neutrino-nucleus scattering: Terrestrial and astrophysical applications,'' arXiv:2203.07361 [hep-ph].
\bibitem{NuTeV:2001whx} G.~P.~Zeller \textit{et al.} [NuTeV], ``A Precise Determination of Electroweak Parameters in Neutrino Nucleon Scattering,'' Phys. Rev. Lett. \textbf{88}, 091802 (2002) [erratum: Phys. Rev. Lett. \textbf{90}, 239902 (2003)].
\bibitem{Qweak:2018tjf} D.~Androi\'c \textit{et al.} [Qweak], ``Precision measurement of the weak charge of the proton,'' Nature \textbf{557}, no.7704, 207-211 (2018).
\bibitem{MOLLER:2014iki} J.~Benesch \textit{et al.} [MOLLER], ``The MOLLER Experiment: An Ultra-Precise Measurement of the Weak Mixing Angle Using M{\textbackslash{}o}ller Scattering,'' arXiv:1411.4088 [nucl-ex].
\bibitem{Roberts:2014bka} B.~M.~Roberts, V.~A.~Dzuba and V.~V.~Flambaum, ``Parity and Time-Reversal Violation in Atomic Systems,'' Ann. Rev. Nucl. Part. Sci. \textbf{65}, 63-86 (2015).
\bibitem{AristizabalSierra:2017joc} D.~Aristizabal Sierra, N.~Rojas and M.~H.~G.~Tytgat, ``Neutrino non-standard interactions and dark matter searches with multi-ton scale detectors,'' JHEP \textbf{03}, 197 (2018).
\bibitem{Giunti:2019xpr} C.~Giunti, ``General COHERENT constraints on neutrino nonstandard interactions,'' Phys. Rev. D \textbf{101}, no.3, 035039 (2020).
\bibitem{Liao:2017uzy} J.~Liao and D.~Marfatia, ``COHERENT constraints on nonstandard neutrino interactions,'' Phys. Lett. B \textbf{775}, 54-57 (2017).
\bibitem{Dent:2017mpr} J.~B.~Dent, B.~Dutta, S.~Liao, J.~L.~Newstead, L.~E.~Strigari and J.~W.~Walker, ``Accelerator and reactor complementarity in coherent neutrino-nucleus scattering,'' Phys. Rev. D \textbf{97}, no.3, 035009 (2018).
\bibitem{Dent:2019ueq} J.~B.~Dent, B.~Dutta, D.~Kim, S.~Liao, R.~Mahapatra, K.~Sinha and A.~Thompson, ``New Directions for Axion Searches via Scattering at Reactor Neutrino Experiments,'' Phys. Rev. Lett. \textbf{124}, no.21, 211804 (2020).
\bibitem{Miranda:2020syh} O.~G.~Miranda, D.~K.~Papoulias, O.~Sanders, M.~T\'ortola and J.~W.~F.~Valle, ``Future CEvNS experiments as probes of lepton unitarity and light-sterile neutrinos,'' Phys. Rev. D \textbf{102}, 113014 (2020).
\bibitem{Suliga:2020jfa} A.~M.~Suliga and I.~Tamborra, ``Astrophysical constraints on nonstandard coherent neutrino-nucleus scattering,'' Phys. Rev. D \textbf{103} (2021) no.8, 083002.
\bibitem{Dutta:2020che} B.~Dutta, R.~F.~Lang, S.~Liao, S.~Sinha, L.~Strigari and A.~Thompson, ``A global analysis strategy to resolve neutrino NSI degeneracies with scattering and oscillation data,'' JHEP \textbf{09} (2020), 106.
\bibitem{Dutta:2019nbn} B.~Dutta, D.~Kim, S.~Liao, J.~C.~Park, S.~Shin and L.~E.~Strigari, ``Dark matter signals from timing spectra at neutrino experiments,'' Phys. Rev. Lett. \textbf{124}, no.12, 121802 (2020).
\bibitem{Wolfenstein:1977ue} L.~Wolfenstein, ``Neutrino Oscillations in Matter,'' Phys. Rev. D \textbf{17}, 2369-2374 (1978).
\bibitem{Proceedings:2019qno} P.~S.~Bhupal Dev, K.~S.~Babu, P.~B.~Denton, P.~A.~N.~Machado, C.~A.~Arg\"uelles, J.~L.~Barrow, S.~S.~Chatterjee, M.~C.~Chen, A.~de Gouv\^ea and B.~Dutta \textit{et al.}, ``Neutrino Non-Standard Interactions: A Status Report,'' SciPost Phys. Proc. \textbf{2}, 001 (2019).
\bibitem{Coloma:2016gei} P.~Coloma and T.~Schwetz, ``Generalized mass ordering degeneracy in neutrino oscillation experiments,'' Phys. Rev. D \textbf{94}, no.5, 055005 (2016) [erratum: Phys. Rev. D \textbf{95}, no.7, 079903 (2017)].
\bibitem{Coloma:2017egw} P.~Coloma, P.~B.~Denton, M.~C.~Gonzalez-Garcia, M.~Maltoni and T.~Schwetz, ``Curtailing the Dark Side in Non-Standard Neutrino Interactions,''
JHEP \textbf{04}, 116 (2017).
\bibitem{ParticleDataGroup:2020ssz} P.~A.~Zyla \textit{et al.} [Particle Data Group], ``Review of Particle Physics,'' PTEP \textbf{2020}, no.8, 083C01 (2020).
\bibitem{Balantekin:2013sda} A.~B.~Balantekin and N.~Vassh,``Magnetic moments of active and sterile neutrinos,'' Phys. Rev. D \textbf{89}, no.7, 073013 (2014).
\bibitem{Vogel:1989iv} P.~Vogel and J.~Engel, ``Neutrino Electromagnetic Form-Factors,'' Phys. Rev. D \textbf{39}, 3378 (1989).
\bibitem{Bell:2006wi} N.~F.~Bell, M.~Gorchtein, M.~J.~Ramsey-Musolf, P.~Vogel and P.~Wang, ``Model independent bounds on magnetic moments of Majorana neutrinos,''
Phys. Lett. B \textbf{642}, 377-383 (2006).
\bibitem{Borexino:2017fbd} M.~Agostini \textit{et al.} [Borexino], ``Limiting neutrino magnetic moments with Borexino Phase-II solar neutrino data,''
Phys. Rev. D \textbf{96}, no.9, 091103 (2017).
\bibitem{LSND:2001akn} L.~B.~Auerbach \textit{et al.} [LSND], ``Measurement of electron - neutrino - electron elastic scattering,'' Phys. Rev. D \textbf{63}, 112001 (2001).
\bibitem{Kosmas:2015sqa} T.~S.~Kosmas, O.~G.~Miranda, D.~K.~Papoulias, M.~Tortola and J.~W.~F.~Valle, ``Probing neutrino magnetic moments at the Spallation Neutron Source facility,'' Phys. Rev. D \textbf{92}, no.1, 013011 (2015).
\bibitem{Dodd:1991ni} A.~C.~Dodd, E.~Papageorgiu and S.~Ranfone, ``The Effect of a neutrino magnetic moment on nuclear excitation processes,'' Phys. Lett. B \textbf{266}, 434-438 (1991).
\bibitem{Anderson:2012pn} A.~J.~Anderson, J.~M.~Conrad, E.~Figueroa-Feliciano, C.~Ignarra, G.~Karagiorgi, K.~Scholberg, M.~H.~Shaevitz and J.~Spitz, ``Measuring Active-to-Sterile Neutrino Oscillations with Neutral Current Coherent Neutrino-Nucleus Scattering,'' Phys. Rev. D \textbf{86}, 013004 (2012).
\bibitem{Formaggio:2011jt} J.~A.~Formaggio, E.~Figueroa-Feliciano and A.~J.~Anderson, ``Sterile Neutrinos, Coherent Scattering and Oscillometry Measurements with Low-temperature Bolometers,'' Phys. Rev. D \textbf{85}, 013009 (2012).
\bibitem{Kosmas:2017zbh} T.~S.~Kosmas, D.~K.~Papoulias, M.~Tortola and J.~W.~F.~Valle, ``Probing light sterile neutrino signatures at reactor and Spallation Neutron Source neutrino experiments,'' Phys. Rev. D \textbf{96}, no.6, 063013 (2017).
\bibitem{Blanco:2019vyp} C.~Blanco, D.~Hooper and P.~Machado, ``Constraining Sterile Neutrino Interpretations of the LSND and MiniBooNE Anomalies with Coherent Neutrino Scattering Experiments,'' Phys. Rev. D \textbf{101}, no.7, 075051 (2020).
\bibitem{deNiverville:2015mwa} P.~deNiverville, M.~Pospelov and A.~Ritz, ``Light new physics in coherent neutrino-nucleus scattering experiments,'' Phys. Rev. D \textbf{92}, no.9, 095005 (2015).
\bibitem{deNiverville:2016rqh} P.~deNiverville, C.~Y.~Chen, M.~Pospelov and A.~Ritz, ``Light dark matter in neutrino beams: production modelling and scattering signatures at MiniBooNE, T2K and SHiP,'' Phys. Rev. D \textbf{95}, no.3, 035006 (2017).
\bibitem{Dutta:2022tav} B.~Dutta, W.~C.~Huang, J.~L.~Newstead and V.~Pandey, ``Inelastic nuclear scattering from neutrinos and dark matter,'' Phys. Rev. D \textbf{106}, no.11, 113006 (2022).
\bibitem{Dutta:2023fij} B.~Dutta, W.~C.~Huang and J.~L.~Newstead, ``Probing the dark sector with nuclear transition photons,'' arXiv:2302.10250 [hep-ph].
\end{thebibliography}
\end{document}